\documentclass[12pt]{article}
\usepackage{epsfig}
\usepackage{amsmath}
\usepackage{hhline}
\usepackage{amssymb}
\usepackage{times}

\newlength{\dinwidth}
\newlength{\dinmargin}
\begin{document}  
% The rest
\newcommand{\pom}{{I\!\!P}}
\newcommand{\slowpi}{\pi_{\mathit{slow}}}
\newcommand{\fiidiii}{F_2^{D(3)}}
\newcommand{\fiidiiiarg}{\fiidiii\,(\beta,\,Q^2,\,x)}
\newcommand{\n}{1.19\pm 0.06 (stat.) \pm0.07 (syst.)}
\newcommand{\nz}{1.30\pm 0.08 (stat.)^{+0.08}_{-0.14} (syst.)}
\newcommand{\fiidiiiful}{F_2^{D(4)}\,(\beta,\,Q^2,\,x,\,t)}
\newcommand{\fiipom}{\tilde F_2^D}
\newcommand{\ALPHA}{1.10\pm0.03 (stat.) \pm0.04 (syst.)}
\newcommand{\ALPHAZ}{1.15\pm0.04 (stat.)^{+0.04}_{-0.07} (syst.)}
\newcommand{\fiipomarg}{\fiipom\,(\beta,\,Q^2)}
\newcommand{\pomflux}{f_{\pom / p}}
\newcommand{\nxpom}{1.19\pm 0.06 (stat.) \pm0.07 (syst.)}
\newcommand {\gapprox}
   {\raisebox{-0.7ex}{$\stackrel {\textstyle>}{\sim}$}}
\newcommand {\lapprox}
   {\raisebox{-0.7ex}{$\stackrel {\textstyle<}{\sim}$}}
\def\gsim{\,\lower.25ex\hbox{$\scriptstyle\sim$}\kern-1.30ex%
\raise 0.55ex\hbox{$\scriptstyle >$}\,}
\def\lsim{\,\lower.25ex\hbox{$\scriptstyle\sim$}\kern-1.30ex%
\raise 0.55ex\hbox{$\scriptstyle <$}\,}
\newcommand{\pomfluxarg}{f_{\pom / p}\,(x_\pom)}
\newcommand{\dsf}{\mbox{$F_2^{D(3)}$}}
\newcommand{\dsfva}{\mbox{$F_2^{D(3)}(\beta,Q^2,x_{I\!\!P})$}}
\newcommand{\dsfvb}{\mbox{$F_2^{D(3)}(\beta,Q^2,x)$}}
\newcommand{\dsfpom}{$F_2^{I\!\!P}$}
\newcommand{\gap}{\stackrel{>}{\sim}}
\newcommand{\lap}{\stackrel{<}{\sim}}
\newcommand{\fem}{$F_2^{em}$}
\newcommand{\tsnmp}{$\tilde{\sigma}_{NC}(e^{\mp})$}
\newcommand{\tsnm}{$\tilde{\sigma}_{NC}(e^-)$}
\newcommand{\tsnp}{$\tilde{\sigma}_{NC}(e^+)$}
\newcommand{\st}{$\star$}
\newcommand{\sst}{$\star \star$}
\newcommand{\ssst}{$\star \star \star$}
\newcommand{\sssst}{$\star \star \star \star$}
\newcommand{\tw}{\theta_W}
\newcommand{\sw}{\sin{\theta_W}}
\newcommand{\cw}{\cos{\theta_W}}
\newcommand{\sww}{\sin^2{\theta_W}}
\newcommand{\cww}{\cos^2{\theta_W}}
\newcommand{\trm}{m_{\perp}}
\newcommand{\trp}{p_{\perp}}
\newcommand{\trmm}{m_{\perp}^2}
\newcommand{\trpp}{p_{\perp}^2}
\newcommand{\alp}{\alpha_s}

\newcommand{\alps}{\alpha_s}
\newcommand{\sqrts}{$\sqrt{s}$}
\newcommand{\LO}{$O(\alpha_s^0)$}
\newcommand{\Oa}{$O(\alpha_s)$}
\newcommand{\Oaa}{$O(\alpha_s^2)$}
\newcommand{\PT}{p_{\perp}}
\newcommand{\JPSI}{J/\psi}
\newcommand{\sh}{\hat{s}}
\newcommand{\uh}{\hat{u}}
\newcommand{\MP}{m_{J/\psi}}
\newcommand{\PO}{I\!\!P}
\newcommand{\xbj}{x}
\newcommand{\xpom}{x_{\PO}}
\newcommand{\ttbs}{\char'134}
\newcommand{\xpomlo}{3\times10^{-4}}  
\newcommand{\xpomup}{0.05}  
\newcommand{\dgr}{^\circ}
\newcommand{\pbarnt}{\,\mbox{{\rm pb$^{-1}$}}}
\newcommand{\gev}{\,\mbox{GeV}}
\newcommand{\WBoson}{\mbox{$W$}}
\newcommand{\fbarn}{\,\mbox{{\rm fb}}}
\newcommand{\fbarnt}{\,\mbox{{\rm fb$^{-1}$}}}
%
% Some useful tex commands
%
\newcommand{\qsq}{${Q^2}$ }
\newcommand{\gevsq}{${\mathrm{GeV}^2}$ }
\newcommand{\etd}{{\mbox{$E_T^*$}} }
\newcommand{\rap}{${\eta^*}$ }
\newcommand{\gp}{${\gamma^*}p$ }
\newcommand{\dsiget}{${{\rm d}\sigma_{ep}/{\rm d}{\mbox{$\langle E_T^* \rangle$}}}$ }
\newcommand{\dsigrap}{${{\rm d}\sigma_{ep}/{\rm d}\eta^*}$ }
\def\Journal#1#2#3#4{{#1} {#2} (#4) #3}
\def\NCA{ Nuovo Cimento}
\def\NIM{ Nucl. Instr. and Meth.}
\def\NIMA{{ Nucl. Instr. and Meth.} A}
\def\NPB{{ Nucl. Phys.} B}
\def\NP{{ Nucl. Phys.}}
\def\PLB{{ Phys. Lett.}  B}
\def\PRL{ Phys. Rev. Lett.}
\def\PRD{{ Phys. Rev.} D}
\def\ZPC{{ Z. Phys.} C}
\def\EPC{{ Eur. Phys. J.} C}
\def\PR{{ Phys. Rev. }}
\def\IJMPA{{ Int. J. Mod. Phys. } A}
% here is an example how to use it :
% \bibitem{ellis}R.K. Ellis, and P. Nason, \Journal{\NPB}{312}{551}{1989}.

\begin{titlepage}
\begin{flushleft}
{\tt DESY 99-091    \hfill    ISSN 0418-9833 \\
  July 1999}                  \\
\end{flushleft}

\begin{flushleft}
%{\bf Draft 2.0} \\
%{\bf Editors: }M.~Kapishin, D.~Kr\"ucker and D.~Milstead \\
%{\bf Referees: }M.~Fleischer and T.~Greenshaw \\
%{\bf Comments to the Editors and Referees by:} 04/06/99 12:00 \\  
\end{flushleft}

\vspace{3cm}

\begin{center}
\begin{Large}

{\bf Measurements of Transverse Energy Flow in Deep-Inelastic Scattering
at HERA}

\vspace{2cm}

H1 Collaboration

\end{Large}
\end{center}

\vspace{2cm}

\begin{abstract}
 \noindent Measurements of transverse energy flow are
presented for neutral current deep-inelastic
 scattering events produced in positron-proton collisions at HERA. The
 kinematic range covers squared momentum transfers $Q^2$ from
 $3.2$ to $2\,200$ GeV$^2$,  the Bjorken scaling variable $x$
 from $8\cdot10^{-5}$ to $0.11$ and  the hadronic 
mass $W$ from $66$ to $233$ GeV. The 
 transverse  energy flow is measured in the hadronic centre of mass
 frame and is studied as a function of $Q^2$, $x$, $W$ and
pseudorapidity.  
  A comparison is made with QCD based 
  models.
 The behaviour of the mean transverse energy 
in the central 
 pseudorapidity region 
  and an interval corresponding to the 
 photon fragmentation region are analysed as a
function of  $Q^2$ and $W$.
% $x$, $Q^2$ and as a function of t $W$, of the hadronic final state.

\end{abstract}

\vspace{2.cm}
\hspace{5.0cm} {\it To be submitted to Eur. Phys. J. C }

\end{titlepage}

\begin{flushleft}
%   H1AUTS  Author list by names, no. of authors  360
%           status: 15/01/99   11.22.57
 C.~Adloff$^{33}$,                %WUPP-ST                  Adloff             
 V.~Andreev$^{24}$,               %LPI -PD                  Andreev            
 B.~Andrieu$^{27}$,               %ECPL-PD                  Andrieu            
 V.~Arkadov$^{34}$,               %ZEUT-ST    10/96         Arkadov            
 A.~Astvatsatourov$^{34}$,        %ZEUT-ST     2/98         Astvatsatourov     
 I.~Ayyaz$^{28}$,                 %PARI-ST       5/96       Ayyaz              
 A.~Babaev$^{23}$,                %ITEP-PD                  Babaev             
 J.~B\"ahr$^{34}$,                %ZEUT-PD                  Baehr              
 P.~Baranov$^{24}$,               %LPI -PD                  Baranov            
 E.~Barrelet$^{28}$
% ...

\end{flushleft}
\begin{flushleft} {\it

%     H1 Institutes as appearing on publications
% ...
$ ^2$ III. Physikalisches Institut der RWTH, Aachen, Germany$^a$ \\
% ...
$ ^{23}$ Institute for Theoretical and Experimental Physics,
          Moscow, Russia \\
 $ ^{24}$ Lebedev Physical Institute, Moscow, Russia$^{f,k}$ \\
% ...
$ ^{27}$ LPNHE, \'{E}cole Polytechnique, IN2P3-CNRS, Palaiseau, France \\
 $ ^{28}$ LPNHE, Universit\'{e}s Paris VI and VII, IN2P3-CNRS,
          Paris, France \\
% ...
$ ^{33}$ Fachbereich Physik, Bergische Universit\"at Gesamthochschule
          Wuppertal, Wuppertal, Germany$^a$ \\
 $ ^{34}$ DESY, Zeuthen, Germany$^a$ \\
% ...
$ ^a$ Supported by the Bundesministerium f\"ur Bildung, Wissenschaft,
        Forschung und Technologie, FRG,
        under contract numbers 7AC17P, 7AC47P, 7DO55P, 7HH17I, 7HH27P,
        7HD17P, 7HD27P, 7KI17I, 6MP17I and 7WT87P \\
% ...
$ ^f$ Supported by the Deutsche Forschungsgemeinschaft \\
% ...
$ ^k$ Supported by Russian Foundation for Basic Research 
       grant no. 96-02-00019 \\
% ...

  } \end{flushleft}

\newpage

%\thispagestyle{empty}\phantom{jj}\clearpage\setcounter{page}{1}

% dk
\def\CPC{Comp. Phys. Comm.}
\def\qq{{$Q^2$} }
\def\etf{$E_T$ flow~}
\def\te{transverse energy~}
\newcommand\mean[1]{\mbox{$\langle #1 \rangle$}}
\section{Introduction}  
Measurements of hadronic final state quantities 
are extremely useful in investigating the different QCD processes that
occur in the wide range of phase space made accessible by the 
 $ep$ collider HERA.
% is a unique facility
%that has opened up a new kinematic domain in the study of $ep$
%scattering.
%It has so far enabled  studies to be made  
%for values of $x$, the Bjorken scaling variable, 
%of between $3 \cdot 10^{-5}$ and $0.85$ and $Q^2$, the virtuality of the 
%exchanged boson, of up to $50\, 000$ GeV$^2$. 
 One such quantity is transverse energy,
measurements of 
which contain global information about charged and neutral particles
and cover a wider pseudorapidity range than  equivalent available charged 
track analyses~\cite{ua1,charged,mk1}.

Electron-proton scattering for values of $Q^2$, the virtuality of the 
exchanged boson,  
significantly above $1\,$
GeV$^2$ is 
usually considered 
as a deep inelastic scattering (DIS) process, in which an exchanged  
 boson directly couples to a parton in the proton. 
This approach successfully describes inclusive  
cross-section measurements~\cite{h1f297,zeusf297},
provided that 
appropriate parton distribution functions are used to describe the partonic 
content of the proton. Several approaches, based  on the
DGLAP~\cite{DGLAP}, 
BFKL~\cite{bfkl} and CCFM~\cite{CCFM} equations are available for  
the QCD evolution of these parton densities to an appropriate scale before 
the interaction with the exchanged boson.   

Extending this picture to describe the hadronic final state  
introduces a number of further complications. It becomes not only 
necessary to understand what happens to the parton involved in the 
partonic scattering process in more detail, but also to understand the 
effects of the interaction on the entire proton. Thus the influence of 
the evolution process leading to the parton undergoing the hard 
scattering must be modelled, as must the behaviour of the proton 
remnant and the fragmentation process. 
%The models rely on various 
%schemes for generating showers of partons from those involved in the 
%primary interaction and are based on the QCD evolution equations 
%%or phenomenological approaches, such as  the Colour 
%%Dipole Model~\cite{dipole}.
%such as DGLAP based prescriptions, 
%the colour dipole model, the BFKL prescription and the linked dipole 
%chain.
%% Phenomenological schemes are also used to attempt to describe the 
%%behaviour of the proton remnant and to provide 
%%a prescription for converting the produced partons into hadrons. 
%One of the aims of this paper is to confront 
%available models of these processes
%based on these ideas 
%with the data and identify 
%those areas where they are successful and those where they are deficient.  
Measurements of transverse energy flow~\cite{mk1,mk2} have proven useful 
in discriminating between the different approaches used 
in these QCD models. 
%  Earlier studies~\cite{mk1,mk2} have shown that
%transverse energy flow, relative to the photon-proton axis in the
%hadronic centre of mass system,  
%is sensitive  to the modeling of the hadronic final 
%state. 
For example, it was shown that early Monte Carlo models based on
DGLAP evolution tended to produce insufficient transverse energy in the
region near the proton remnant for values of the Bjorken 
scaling variable $x$ of less than about $10^{-3}$. 
% a defect not present %in predictions made using the Colour Dipole
%Model. 
  
Recent measurements of jet and leading particle production in 
DIS~\cite{tania,zeusfj,dave} suggest that   
the description of the data provided by DGLAP based models 
can be improved by allowing the virtual photon to have structure. That
is, in addition to ``direct photon'' events, in which the entire momentum
of the virtual photon enters the hard scattering process, ``resolved
photon'' events are allowed. In these, the virtual photon is considered
to have a developed partonic structure and one of the partons in the 
photon participates in the hard interaction. The significance of this 
idea for the description of the transverse energy flow, in the framework 
of the above models of DIS, is also investigated in this paper.

%As an alternative to the above, it is instructive
%to view
The deep inelastic $ep$ scattering may also be viewed in the rest frame of
the proton. In this frame
  the photon can be considered to fluctuate into a hadronic object which
subsequently interacts with the proton, 
even for $Q^2$ values of up to $1\,000$ GeV$^2$~\cite{bjoko,duca,shaw}.
In the proton rest frame and for values of $Q^2$ above several GeV$^2$  
the fluctuation time~\cite{ioffe,duca} 
is given  by $\tau \approx 1/(xM_p)$, where $M_p$ is the mass of the hadronic target.
For values of  $x$ of less than $10^{-2}$ the virtual photon can fluctuate into 
 and exist as a hadronic object over a distance of $10$ to $1\,000$ fm,
which 
is far larger than the size of the proton. Under the na\"{\i}ve assumption
that this fluctuation behaves like a 
single hadron then  
$ep$ interactions should resemble hadron-hadron 
scattering.
This approach was tested in an earlier publication~\cite{andre1}, 
in which it was demonstrated that the average transverse energy flow in
the central rapidity region in the hadronic centre of mass system showed
no significant 
$Q^2$ dependence. This is consistent with observations  made in
hadron-hadron scattering, in which particle production in the central 
region is largely independent
of the nature of the hadronic object and depends only
on the total centre of mass energy~\cite{na22}. Conversely, 
transverse energy production in the  expected photon fragmentation region
was found to 
be  strongly dependent on $Q^2$. 
The 
%increased 
data collected by H1 in recent years now allow    
this picture to be studied with far greater precision than was possible in the 
previous work~\cite{andre1}. 

  It is the aim of this analysis to study transverse energy production 
within both the traditional DIS framework and in the picture given
in the proton rest frame. Distributions showing the 
dependence of transverse energy on pseudorapidity, $x$, $Q^2$ and  
the total hadronic mass, $W$,  
are
presented and  quantitative comparisons of QCD based models are made 
with the data. A
more qualitative approach is adopted in examining the
interpretation of our data as 
a hadron-hadron scattering process.  The $Q^2$ and $W$ 
dependence of transverse energy production in the central pseudorapidity
region and a region associated with the fragmenting photon  is
investigated and
compared to hadron-hadron data and earlier H1 results in
both photoproduction and DIS. 

The analysis presented here is based on data taken at 
the HERA collider for which $820$ GeV protons were collided with 
$27.5$ GeV positrons at a centre of mass energy of $300$ GeV. 
Transverse energy production in the hadronic centre of mass system 
is studied in  the kinematic range 
$3.2$ GeV$^2$ $< Q^2<$ $2\,200$ GeV$^2$, 
$x> 8 \cdot 10^{-5}$ and $66<W<233$ GeV. An order of magnitude more data
are
used than
 in the previous H1 measurements~\cite{mk2,andre1}. Furthermore, the phase
space region has been extended into the proton remnant fragmentation
region and now covers more than 8 units of pseudorapidity.

%%% Local Variables: 
%%% mode: latex
%%% TeX-master: "fp"
%%% End: 

\section{The H1 Detector}\label{detector}
A detailed description of the H1 apparatus can be found elsewhere~\cite{h1nim}.
The following section briefly describes the components of the detector relevant
to this analysis. 

The H1 liquid
argon~(LAr) calorimeter~\cite{larc} was used to measure positrons scattered into
the central and forward (proton direction) parts of the H1
detector and also to determine the hadronic energy flow. The
 calorimeter extends over
the polar angle range $4^\circ < \theta <  154^\circ$ with full azimuthal
coverage, where $\theta$ is defined with respect to the incoming proton 
direction. It consists of
an electromagnetic section with lead absorbers and a hadronic section with
steel absorbers. Both sections are highly segmented in the transverse and 
longitudinal directions. 
Energy resolutions for electrons~\cite{h1po} and charged pions~\cite{h1pi}
of
$\sigma_{E_e}/E_e\approx 0.12/\sqrt{E_e\ [\gev]} \oplus 0.01$ and
$\sigma_{E_h}/E_h\approx 0.50/\sqrt{E_h\ [\gev]} \oplus 0.02$,
 respectively, have been 
established in test beams. The uncertainties in the absolute 
electromagnetic and hadronic energy scales are $3$\% and
$4$\%, respectively, for the present data sample.
%~\cite{scale}.

The backward electromagnetic lead-scintillator calorimeter
(BEMC) was used to 
measure the properties of the scattered positron for polar angles in the
range $155^\circ<\theta<176^\circ$. An energy resolution of
$0.10/\sqrt{E_e [\gev]}\oplus 0.42/E_e$[GeV]$\oplus0.03$ 
% Botterweck's paper.
  has been achieved~\cite{h1bemc}. The absolute 
electromagnetic scale has been determined to a precision of
$1$\%~\cite{f2pap}. Since it consists of only one interaction length 
of material, the hadronic response of the BEMC is poor and approximately 
30\% of incident hadrons
%particles 
leave no significant energy deposition. 
Consequently, a large  
scale uncertainty of 20\% exists for hadronic measurements made with this
device. For results presented here based on 1994 data, the BEMC was 
used both to measure properties of the scattered positron and 
the hadronic energy flow. 

The BEMC was replaced in 1995 by the SPACAL  as the 
main rear calorimeter in the H1 detector. This contains  electromagnetic 
and hadronic sections, achieving energy resolutions of $0.075/\sqrt{E_e\ 
[\gev]}\oplus 0.025$
%\cite{SPACALTEST}  
and $30\%$ 
 for positrons and hadrons, 
respectively~\cite{shad}. The absolute electromagnetic energy scale
is known to a precision of 2\% and the hadronic energy scale to 
7\%~\cite{spaca}. For the analysis presented here, 
the SPACAL is used to measure hadronic energy flow for events in which the
scattered positron is reconstructed in the LAr.

A series of tracking chambers are in place which provide the measurement
of the polar angle of the scattered positron in this analysis. Forward and
central tracking chambers 
cover polar angle ranges of $5^\circ<\theta<25^\circ$ and $25^\circ<\theta<155^\circ$,
respectively, and provide a measurement of the primary vertex position and
the polar 
angle of the scattered positron. In the backward region 
($155^\circ<\theta<176^\circ$) the scattered positron is measured by the 
Backward 
Proportional Chamber (BPC) which lies in front of the BEMC. 
%When used with the position of the primary vertex,
These detectors allow the polar angle of the scattered positron
to be measured with a precision of $1$ mrad~\cite{h1f297}.

The \te flow measurement is extended in the proton direction by a small
calorimeter (PLUG) with copper absorber and silicon pad readout, covering the
region between the beam-pipe and the LAr cryostat ($0.7^\circ < \theta <
3.3^\circ$). 
Owing to the large amount of
passive material  between the PLUG and the interaction point, which varies
between $0.8$ and $5$ interaction lengths, less than $40$\% of the energy
measured
in
the PLUG  originates from the primary vertex. Using a full simulation of the 
material distribution in and around the H1 detector, energy loss 
corrections have been 
determined. The precision of these corrections has been studied using  
 $ep$ data~\cite{enrico}.    An energy scale uncertainty of 26\% and 
an energy resolution of $ 1.5/\sqrt{E_h\ [\gev]}$
%$\sigma_{E}/E\approx 0.084$
have been achieved.

\section{Event Selection}\label{select} 
In this analysis three independent event samples are considered: two from
the 1994 and one from the 1996 data taking period. 
The 1994 low \qq samples consist of events in which a scattered 
positron is found in the BEMC and correspond to 
 running periods in which  
 the $ep$ interaction point was at the nominal position and in which 
it was 
shifted by about 67 cm in the proton direction, $z$. This shift allowed 
lower \qq values, down to 2.5 GeV$^2$, to be accessed. 
The 1994 data samples together permit the range
$2.5$~GeV$^2 < Q^2 <  100$ 
GeV$^2$ to be studied.  The integrated luminosities of the nominal and shifted  
vertex data samples  are $2.7$ pb$^{-1}$ and $0.058$ pb$^{-1}$, respectively. 
%In order to reach 
 
 The high \qq sample consists of events 
collected in 1996 and corresponds to an integrated luminosity 
of $8.2$ pb$^{-1}$.  For this 
sample, the scattered positron is detected in the
LAr calorimeter and  \qq is required to be larger  than $100$ GeV$^2$.

DIS events are selected by demanding a well-reconstructed scattered positron
with an energy larger than $12$ GeV.  The event kinematics are determined using
the scattered positron energy, $E_e'$, and polar angle, $\theta_e$ 
 ({electron
method}):
  $ Q^{2} = 4 E_{e} E_{e}^{'}\; {\rm cos}^{2}(\theta_{e} /2)$
and $ y_e = 1 - (E_{e}^{'} / E_{e})\; {\rm sin^2}(\theta_{e} /2)$
 where $E_e$ is the incident positron beam energy.
 The
scaling variable Bjorken-$x$ is related to these quantities via the square
of the centre of mass energy $s$: $x = Q^{2} / (ys)$, and the 
hadronic invariant mass squared is $W^2=sy-Q^2$.

 In addition, the following selection criteria are
applied to suppress non-$ep$ and photoproduction background and to 
maintain
optimal  resolution in the kinematic variables: 
 
\begin{itemize}
\item A reconstructed event vertex must be found within $\pm 30$ cm in 
 $z$ of the nominal interaction point. 
\item The longitudinal momentum balance must be within   $30 \mbox{ GeV} <
      \sum_i E_i - P_{zi} < 70$ GeV, where the sum runs over all energy
      deposits in the LAr and the backward calorimeter.

\item $W_e^2$, the invariant mass squared of the hadronic final state  
determined using the scattered positron 
 is required to be larger 
than $4\,400\,$GeV$^2$.

\item The invariant mass of the hadronic final state determined using 
the hadronic energy deposited in the calorimeters, $W_h$, is 
also required to satisfy $W_h^2 > 4\,400\,$GeV$^2$.

\item To reduce background from photoproduction interactions, $y_e < 0.6$
is required. 
\end{itemize}

For the low \qq sample the positron candidate is required to be found within
the angular acceptance of the BEMC ($157^\circ<\theta_e<173^\circ$ for the
nominal vertex data and
$164^\circ<\theta_e<176^\circ$  for the shifted vertex data). In addition,
positron identification cuts to suppress photoproduction background are made
using the information obtained from the BEMC cluster radius and the matching of
the cluster position with that of a charged particle measured in the BPC. 
%After these cuts the
%remaining photoproduction background is estimated to be less than
%$3$\%~\cite{h1f297}.

For the high \qq sample the scattered positron candidate is required to be
found within the LAr calorimeter ($\theta_e<150^\circ$).  Since the high
\qq
analysis is especially sensitive to QED radiation collinear with the $e^+$ beam,
further selection criteria are included:
\begin{itemize}
\item $0.5 < y_h/y_e < 1.3$. The variable $y_h$ is given by the Jacquet
       Blondel method~\cite{jablo}: $y_h=\frac{1}{2}\Sigma/E_e$ where
       $\Sigma=\sum_i(E_i-p_{zi})$ and $i$ refers to all  energy clusters
       detected in the calorimeters except that due to the scattered positron.
\item In addition to the {electron method}, the $\Sigma$ {method
}~\cite{sigma} is 
        used to calculate $x$ and \qq and a  requirement is made that both 
       methods should prescribe the same kinematic bin for a given event. 
       According to the $\Sigma$ {method} \qq, $y$ and $x$ are given by
       \\
       $Q^2_{\Sigma}={E{_e'}^2} {\sin}^2 \theta_e/(1-y_{\Sigma})$, 
       $y_{\Sigma}=\Sigma /(\Sigma+E_e'(1-\cos \theta_e))$ and
       $x_{\Sigma}=Q_{\Sigma}^2/(s y_{\Sigma})$.
\end{itemize}

After the application of these cuts, the
efficiency for selecting events for every interval in $x$ and $Q^2$ used 
in this study is greater than 75\%. 

%%\noindent The latter conditions are designed to remove events suffering
%%from strong QED initial
%5state radiation or with a badly reconstructed positron. 
%The kinematic plane is divided into several regions in $x$ and
%\qq. The sizes of these bins are adapted to the experimental
%resolutions
% to keep
%migration effects small.
%Careful studies have been performed to prove that these
%cuts do not produce any bias in our results.

%An analysis has also been performed on 
The high \qq data from 1994 have also been studied.
The \te flow measurements made with the two high \qq data sets are
consistent within their statistical errors.
Because of the different correction
procedures necessary for the 1994 and the 1996 data samples and the
significantly 
larger amount of data available to 
 the 1996  high \qq analysis, the two
measurements have not been combined and for the high \qq region
only results based on 1996 data are presented here.

\section{Experimental Method}

%Transverse energy is given by $E_T=E \sin
%\theta$ 
%For a particle with energy $E$ and emission angle $\theta$  with respect
%to the
%incoming
%proton direction, the transverse energy is defined by   $E_T=E \sin
%\theta$.

\begin{figure}
\unitlength1cm
  \begin{picture}(6,5)  \put(0,-4.25){
     \put(0.2,4){\psfig{file=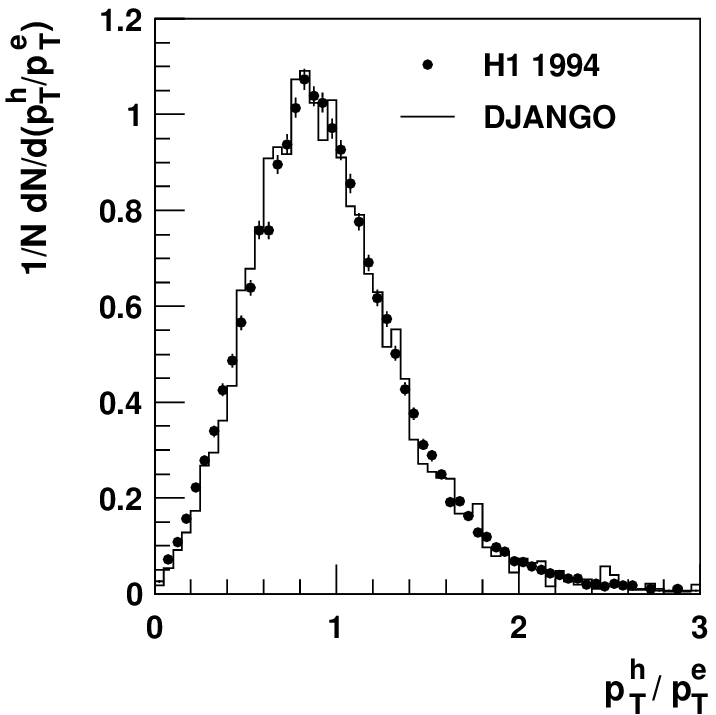,width=5cm,clip=}}
     \put(5.5,4){\psfig{file=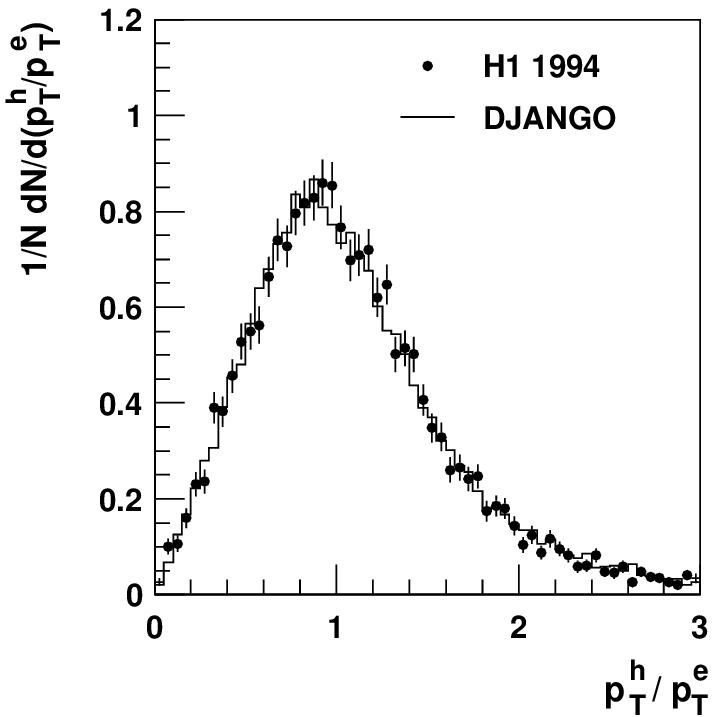,width=5cm,clip=}}
     \put(10.7,4.15){\psfig{file=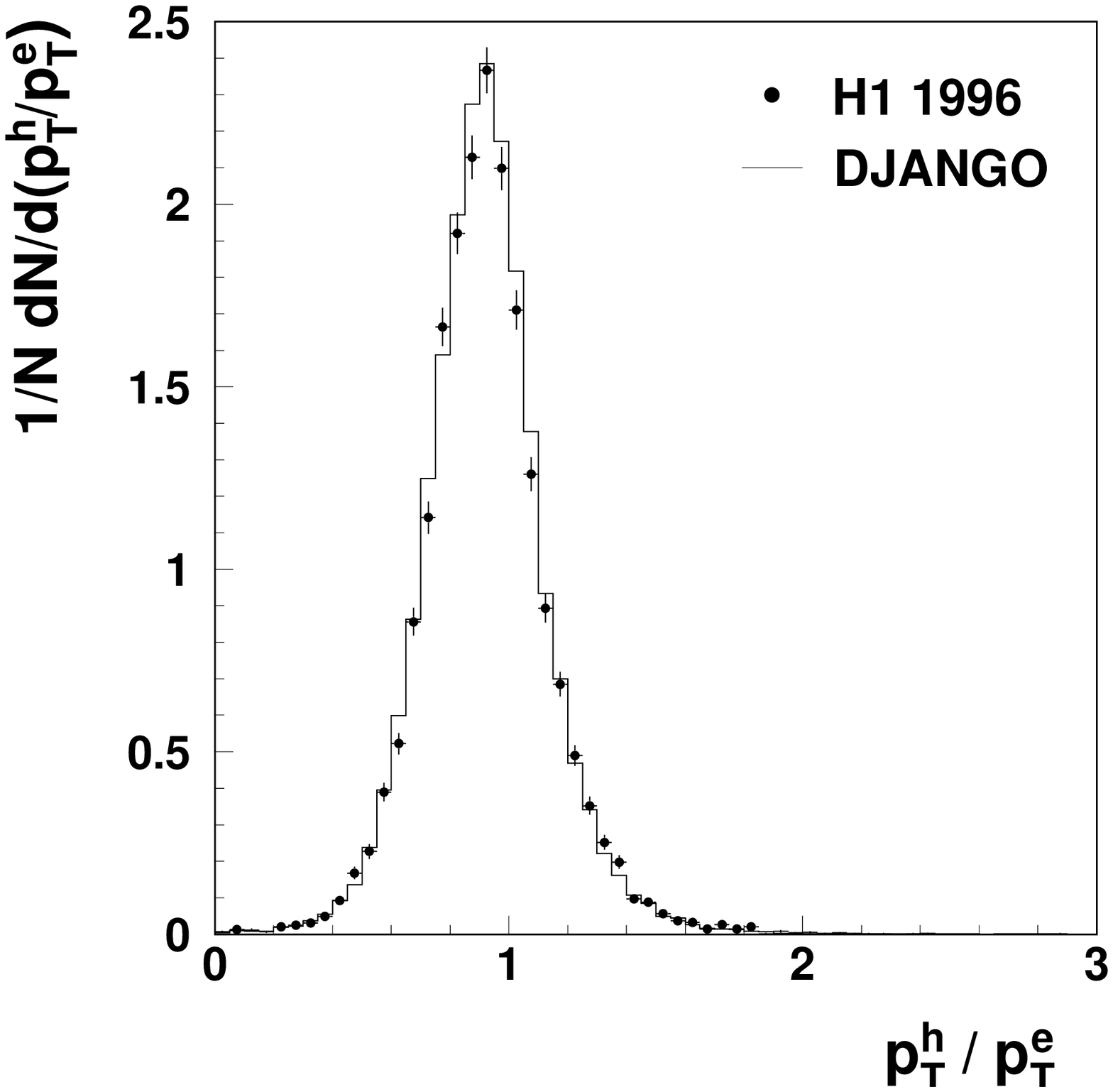,width=5.08cm,clip=}}
                                }
  \end{picture}
  \caption{\label{ptbal} The distribution of $p_T^h/p_T^e$
   measured  with the 1994 nominal vertex data-set
 (left), the 1994 shifted vertex data-set (middle)
and the 1996 data-set (right). The
data are compared with predictions from {\sc Django}.}
\end{figure}

The data are corrected bin-by-bin for QED radiation and detector 
effects using the event generator {\sc Django}~\cite{DJANGO}  
together with a full simulation of the detector response. 
To demonstrate the accuracy of the simulation and the understanding of the
response of the H1 calorimeters the distributions of transverse momentum 
balance, $p_T^h/p_T^e$,  are shown in
Fig.~\ref{ptbal}
 for the different data samples used in this analysis. Here,
$p_T^h$ is defined as the sum of  
the azimuthal four-momentum vector components $x$ and $y$ of each energy 
deposition $i$ in the LAr and SPACAL calorimeters according to
$(p_T^{h})^{2}={(\Sigma}_i 
 p_{x,i})^2 +{(\Sigma}_i p_{y,i})^2$. Similarly, $p_T^e$
is the transverse momentum of the scattered positron. 
 Predictions made with the {\sc Django} model are also 
shown and they describe the data well.

The work presented in this paper comprises studies of the 
production of transverse energy $E_T^*$ as a function of pseudorapidity $\eta^*$
and the kinematics of the $ep$ scattering  process. Measurements are made of   
 the distribution  $1/Nd\etd/d\eta^*$ in which, for a specified kinematic range,     
$N$ is the total number of DIS events and $d\etd/d\eta^*$ is the 
sum of the transverse energies of each particle $i$ per unit of
pseudorapidity. The transverse energy,  $E_{Ti}^*$, of a particle $i$
 with energy $E_i^*$ and polar angle $\theta_i^*$ is defined as
$E_{Ti}^*=E_i^*\sin\theta_i^*$. Results are given in the
hadronic centre of  mass
 system (hCMS) for which the incoming photon direction defines the $+z^*$ 
direction.\footnote{All quantities presented in the hCMS frame are denoted by the superscript *} 

Whereas in the laboratory frame (LAB) a
sizeable contribution to the transverse energy is produced by the
kinematic recoil from the scattered lepton, in the hCMS transverse
energy
is due largely to perturbative QCD and fragmentation effects. The Lorentz
transformation from the LAB into the hCMS therefore  serves to isolate the
physically interesting part of the \etf\hspace{-.85ex}.  

At high \qq the Lorentz transformation from the LAB to the hCMS requires great
care. Any deviation from the true transformation produces an artificial
contribution to the \te seen in the hCMS. This is a more severe problem at high
\qq than at low \qq since the mean particle transverse energy
increases with $Q$ in the LAB but depends 
only weakly on $Q$ in  the
hCMS. For the high \qq data sample the maximum $\langle E_T\rangle$ 
per unit pseudorapidity is 
typically an order of magnitude larger in the LAB than in the
hCMS (about 20~GeV compared to 2~GeV). In practice the transformation is
calculated from the energy and direction of the scattered positron. The 
main sources of error are the energy resolution and
calibration of the calorimeters and QED radiation.  Although the energy
resolution and calibration of the H1 LAr
calorimeter are well understood and the influence of QED radiation  is well
described by {\sc Django}, these remain an important source of systematic
bias. 
To minimise their influence on the measurement of the transverse energy
for the high \qq data,  the method described below  was used. 
This exploits the precise information which is available on the direction
of the scattered
lepton.

A deviation in the energy measurement of the scattered positron  results  
in 
an artificial component in the momentum of any particle in the hCMS which
lies within the scattering plane
defined by the incoming proton and the scattered positron. To suppress  
this artificial momentum, the variable used to measure \te 
in the high \qq data set is redefined as 
$E_{\perp}^*=E_T^* |\sin \varphi^*|\frac{\pi}{2}$, where $\varphi^*$
is the  %polar angle of a hadron w.r.t.\ to the scattered positron.
 azimuthal angle w.r.t.\ the lepton scattering plane.  
Integrated over
$\varphi^*$ this is again equal to $E_T^*$, assuming an isotropic
distribution of the true \te around the proton in the hCMS in $\varphi^*$.
Therefore the
measurements of 
\mean{E_T^*} and \mean{E_{\perp}^*} are equivalent for an isotropic $\varphi^*$
distribution.
Although $\varphi^*$
asymmetries are predicted in pQCD from processes such as boson-gluon fusion 
and QCD-Compton scattering they  introduce biases of less than 1\% 
in the transverse energy flow at high $Q^2$. This is 
%anyway 
a significantly 
smaller effect than the biases due to poor \etd resolution which would be
introduced were the redefinition not used.   

\section{Systematic Errors}

There are several sources of systematic effects which may 
affect the measurements presented in this paper. The systematic  
errors owing to these have been investigated~\cite{enrico,fabian} and are
discussed below. 

%different errors are added in quadrature and have been estimated for each
%region in $x$, \qq and $\eta^*$ separately.

%The effects of the following sources of systematic
%uncertainties on the transverse energy measurements have been
%studied~\cite{enrico,fabian} for the distributions presented in this
%paper:
\begin{itemize}
   \item The hadronic energy scales of the calorimeters are known to an
         accuracy of $4$\% (LAr), $26$\% (PLUG), $7$\% (SPACAL) and $20$\%
         (BEMC). This directly gives the uncertainties for the energy
         measurements presented here.
   \item For the scattered positron the energy calibration uncertainty is 
        $1$\% in the BEMC region (low \qq sample) and  $3$\% for the
         LAr calorimeter (high \qq sample). Resulting errors on  
         measured {\etd} are typically
         $2$\% at low \qq  and $6$\% at high \qq. At high \qq and at large
         values of  $\eta^*$ this error can become large ($\le 27\%$).  
         Section \ref{tef} describes in more detail the specific
         problems related to the high \qq
         measurements. 
   \item Uncertainties owing to the reconstruction of the kinematic
         variables $x$ and 
         \qq are typically of the order of $2$\% and $1.5\%$ for the low
         \qq  and high \qq spectra, respectively. The improved precision at high \qq 
         is due to the additional
         cuts applied to this data sample (see section \ref{select}).
   \item The model dependence of the bin-by-bin correction is estimated 
         using the event generators   
         {\sc Ariadne}~\cite{ariadne}, 
         {\sc Herwig}~\cite{herw} and {\sc Lepto}~\cite{lepto}  with two
parton density  functions (GRV~\cite{GRV} and MRSH~\cite{MRSH}). 
         The differences between the models are typically $3\%$ at 
high \qq  and
$4.5$\% at low
         $Q^2$. Larger differences are found for the two $\eta^*$ bins measured
         with the PLUG calorimeter ($20$\%). The larger
 errors in the  PLUG pseudorapidity region are due to the uncertainties 
in the modelling of the physics in this region. 
   \item The \te measurement is strongly reliant on a correct simulation
of     the inactive material in and around the H1 detector, particularly 
         in the forward region.   
          The sensitivity of the measurements to the 
         assumed material distribution is estimated by varying the 
         amount and location of the inactive material in the simulation. 
         The transverse energy flows measured with the 1994 and 1996 
         data, for which different configurations of dead material 
          were present, are also compared.  
         This gives rise to  typical  uncertainties 
         of the order of $4\%$ for the  
data samples. These differences are
included as systematic errors. For the 
         points measured with the PLUG 
         calorimeter a different procedure is used. Using the shifted
vertex  and the nominal vertex data sets energy flow is measured 
in approximately the same detector volume in the PLUG although 
incident particles pass through different amounts of inactive material.
%is different 
 After applying the dead material 
correction,  the change in total 
energy measured in the PLUG  is
compatible with Monte Carlo  expectations to within $5$\%. The magnitude
of
the 
total energy shift observed in data 
between the nominal and shifted vertex event samples, $11$\%, is taken as
 the systematic error on this measurement due to the influence of
 dead material. 
   \item Another possible source of error is the simulation of the hadronic 
         shower within the LAr and the PLUG calorimeters. This has been
         investigated by comparing the simulation programs {\sc
Gheisha}~\cite{geisha}  and {\sc Calor}~\cite{calor}. The differences are
         small for the LAr calorimeter ($3\%$) but large for the PLUG 
         calorimeter  ($20$\%).
   \item Photoproduction background is only important at low $Q^2$. 
         The typical uncertainty on the measured points owing to this
source 
         is  $1\%$ and has been estimated with the
         {\sc Phojet}~\cite{phojet} program.
\end{itemize}

\section{QCD-based Models}\label{QCDpre}

Although progress has been made on several other 
fronts~\cite{evs,mlla,lattice,neweth} it is still the case that most QCD
based
predictions of the hadronic final state are produced with Monte Carlo 
methods which use  the following prescription. Phenomenologically derived
parton distribution functions, evolved to the relevant scale, are used to
determine the properties of the partons  emerging from the initial state
hadron, or photon. Some of these partons 
undergo a hard scatter, the cross-section for which 
is calculated using leading order (LO) QCD. The resulting partons 
undergo fragmentation to produce the observed hadrons.

The proton parton distribution functions are reasonably well constrained 
by inclusive measurements of DIS (in particular $F_2$), and data from
hadron-hadron scattering experiments. The effects 
on the hadronic final state of changing the 
input parton distribution functions within the limits allowed by
the inclusive DIS data are small compared to the effects arising from 
varying other aspects of the calculations. The proton parton 
distribution functions used here are the CTEQ4L~\cite{CTEQ,pdflib} 
set. For calculations involving resolved virtual photon processes, 
 the SAS-G virtual photon parton parameterisation~\cite{sasgam} is 
used.

Different QCD evolution equations are known. They
have been derived in the
Leading-Log-Approximation and  are expected to be valid in certain regions
of Bjorken $x$. The DGLAP evolution 
equation effectively resums the leading
$\log Q^2$ terms which corresponds to the
strong ordering in transverse momentum of successive
parton emissions and is applicable at large $Q^2$.
The BFKL approach sums leading $\log(1/x)$ 
terms and is expected to become significant at small $x$. 
Successive parton emissions in this approach have strongly decreasing
longitudinal momenta, but are not ordered in transverse momentum.
This latter feature is emulated in parton emissions produced  
by the Colour Dipole Model (CDM)~\cite{dipole} in which
partons are radiated from colour dipoles produced in
the hard interaction~\cite{cdmbfkl}. 
 The CCFM evolution equation forms a bridge between the DGLAP and 
BFKL approaches and resums the leading $\log 1/x$ and $\log Q^2$
terms both for inclusive and non-inclusive quantities.
%are based on an angular ordering prescription 
%in the parton cascade which deals with interference 
%between final and initial state radiation. 
To obtain CCFM based hadronic final state predictions,  the Linked Dipole
Chain~\cite{ldc} model is used here. This is a reformulation of  
the CCFM equation and redefines the separation of initial and final 
state  QCD emission using the CDM. 

 The fragmentation of the produced partons typically involves a 
showering process followed by a hadronisation phase. 
 The models to which the data are compared here use the
string~\cite{string} or cluster~\cite{cluster} 
approach to hadronisation.

% as implemented in
%JETSET 7.4~\cite{jetset74}. 

Although the evolution equations discussed above can be derived within the
picture of a parton cascade, neither the DGLAP nor BFKL
approximations describe the details of the hadronic final state. It is
therefore necessary to reformulate these inclusive equations as an
iterative process, the shower algorithm, to produce  
parton emissions in the Monte Carlo models.\footnote{The CCFM
approach is an iterative process by definition.}
The amount of transverse energy depends on these shower algorithms
and 
%by this 
hence
the 
%measurement 
expected levels
of $E_T$ 
%is 
are
sensitive to 
the
different evolution
schemes.

The Monte Carlo models used here are based on various combinations
of the above ideas. They are described briefly in the following text as 
are any changes to the default settings and modifications of 
the Monte Carlo programs. 
%These were made by the programs' authors,
% largely to provide a better   
%description of these measurements after they 
%were 
%first presented. The models
% used are: 

{\sc Ariadne}~\cite{ariadne} version 4.10. This is an implementation 
of the CDM. In order to generate an increase of the transverse energy with
$Q^2$ the
program has been modified; an additional switch has been
introduced~\cite{arpr} 
and is set to MHAR(151)=2. This changes the phase space
restriction for 
radiation from an extended source. In addition, two control
parameters were changed from their default values. The parameter settings used 
are PARA(10)=1.5 and PARA(15)=0.5~\cite{tune}. These alter the suppression
of 
radiation from the proton remnant and the struck quark, respectively. 
The expectations of this model are marked ``{\sc Ariadne} 4.10 mod'' on 
all figures.

The LDCMC~\cite{ldcmc}, version 1.0. This is based on the Linked Dipole
Chain implementation of CCFM evolution and uses the 
CDM to simulate the effect of higher order emissions.
It is incorporated in the framework of the {\sc Ariadne} package. All
parameters used by both programs are set to their default values,  
with the exception of those listed above. The expectations of this model 
are marked ``LDCMC'' on all figures.  

{\sc Lepto} 6.51~\cite{lepto}. {\sc Lepto} matches exact first order QCD
matrix elements to DGLAP based leading log parton showers. 
It also allows two different methods of non-perturbative rearrangement of
the event colour topology. One way is via ``soft colour interactions''
(SCI)~\cite{SCI}, in which low momentum gluons are exchanged between 
partons in the proton remnant and those which are perturbatively produced 
and this is the default option for this version of {\sc Lepto}. 
However,  this scheme leads to an excess of soft particle production at
high $Q^2$~\cite{mcworkshop}. In a different approach, a string
reinteraction scheme based on a  Generalised Area Law (GAL) is now 
available~\cite{GAL}. This allows interactions between the colour 
strings connecting the final state partons, leading to a reduction 
in the total string area. 
%Within the GAL, the total colour configuration of
%the partonic final state is given by the sum of the areas covered by 
%colour connections between each pair of emitted partons  
%$A_{ij}=(p_i^2+p_j^2)-(m_i^2+m_j^2)$, defined by the momentum $p$ 
%and mass $m$ of partons $i$ and $j$. For a large, multi-parton event  
%different sets of colour configurations among
%the partons are possible and string reinteractions can take
%place  
%to preferentially produce events with small areas. 
To use this option, 
retuned values of parameters relating to the hadronisation and  
 the parton shower schemes have to be
used.\footnote{The following parameter settings are used with 
the GAL string re-interaction option: LST(34)=3, PARJ(42)=0.45,
PARJ(82)=2.0, PYPAR(22)=4.0, and PARL(7)=0.10}
However, as is demonstrated later in this report, the implementation
of this scheme leads to a worsening of the description of
our data.  Because of this,  the GAL option is not used in
comparisons with most of the spectra presented here and model expectations 
which use neither GAL nor SCI are marked ``{\sc Lepto} 6.51 mod'' on
these figures. 

{\sc Rapgap} 2.06/48~\cite{rapgap}. {\sc Rapgap} also matches exact first
order QCD matrix 
elements to DGLAP based leading log parton showers. In addition 
to the direct photon processes simulated by {\sc Lepto}, {\sc Rapgap}  
simulates resolved photon interactions in which the virtual photon is 
assumed to have structure, parameterised using an implementation of the 
SAS-G virtual photon parton distributions. The {\sc Rapgap}  
package allows a choice of renormalisation and 
factorisation scales. For predictions presented here,  these are set to
$p_T^2 + Q^2$, where $p_T$ is the transverse momentum of the partons in
the hard scattering. 
The default value for the cut-off necessary to regulate the 
matrix elements is $p_T^2 > {\rm PT2CUT} = 1.0 \,$GeV$^2$. 
This leads to a predicted DIS cross-section which is larger than
measurements 
for $Q^2 > 200\,$GeV$^2$, so here ${\rm PT2CUT} = 4.0\,$GeV$^2$ is used.
The expectations of this model are marked ``{\sc Rapgap} 2.06/48 dir + res'' on 
all figures. If only direct photon interactions are 
allowed, the results of {\sc Rapgap} are very similar to those of {\sc Lepto}. 

To perform the hadronisation step,  all of the above models use string
fragmentation as implemented in JETSET 7.4~\cite{jetset74}.

{\sc Herwig} 5.9~\cite{herw}.
% is  used in this analysis to estimate
%the model dependence of the correction procedure used  with the low
%$Q^2$ data set. 
 This model is based on leading log parton 
showers with
matrix element corrections~\cite{mecor} and implements a
 cluster hadronisation scheme. The expectations of this model are marked
``{\sc Herwig} 5.9'' where shown.

The {\sc Phojet}~\cite{phojet} Monte Carlo program is 
used to calculate the amount of background due to photoproduction
processes. This generator contains LO QCD matrix elements
for hard
subprocesses, a  parton shower model and a
phenomenological description of  soft processes.

As an alternative to the Monte Carlo based approach,
        analytical BFKL calculations~\cite{etbfkl} 
 are available to 
 predict energy flow in the forward region.  
  These calculations are 
based on asymptotic expressions derived from the BFKL equation
at LO and do not include hadronisation 
effects. The predictions of the calculations
are marked ``BFKL Partons'' 
where shown. 

%%% Local Variables: 
%%% mode: latex
%%% TeX-master: "up"
%%% End: 

\section{Results}

\subsection{Transverse Energy}\label{tef}

In addition to the selection criteria outlined in section \ref{select},  a
further cut on the total energy produced in the angular region  $4.4^\circ\! <
\theta <15^\circ$ is also imposed. The energy in this angular region is
required to be larger than $0.5$ GeV  ($E_{fwd}$ cut). This condition was
introduced for the low \qq analysis  to suppress diffractive-like
events, which are characterised
by an absence of hadronic energy in the 
forward region~\cite{h1r,zr}. It
allows a
comparison with  the predictions of QCD based models which is less
dependent on uncertainties
introduced  by the attempt to model  diffractive processes. 

\begin{figure}[p]\unitlength1cm
  \begin{picture}(16,18)
     \put(0.2,0){\epsfig{file=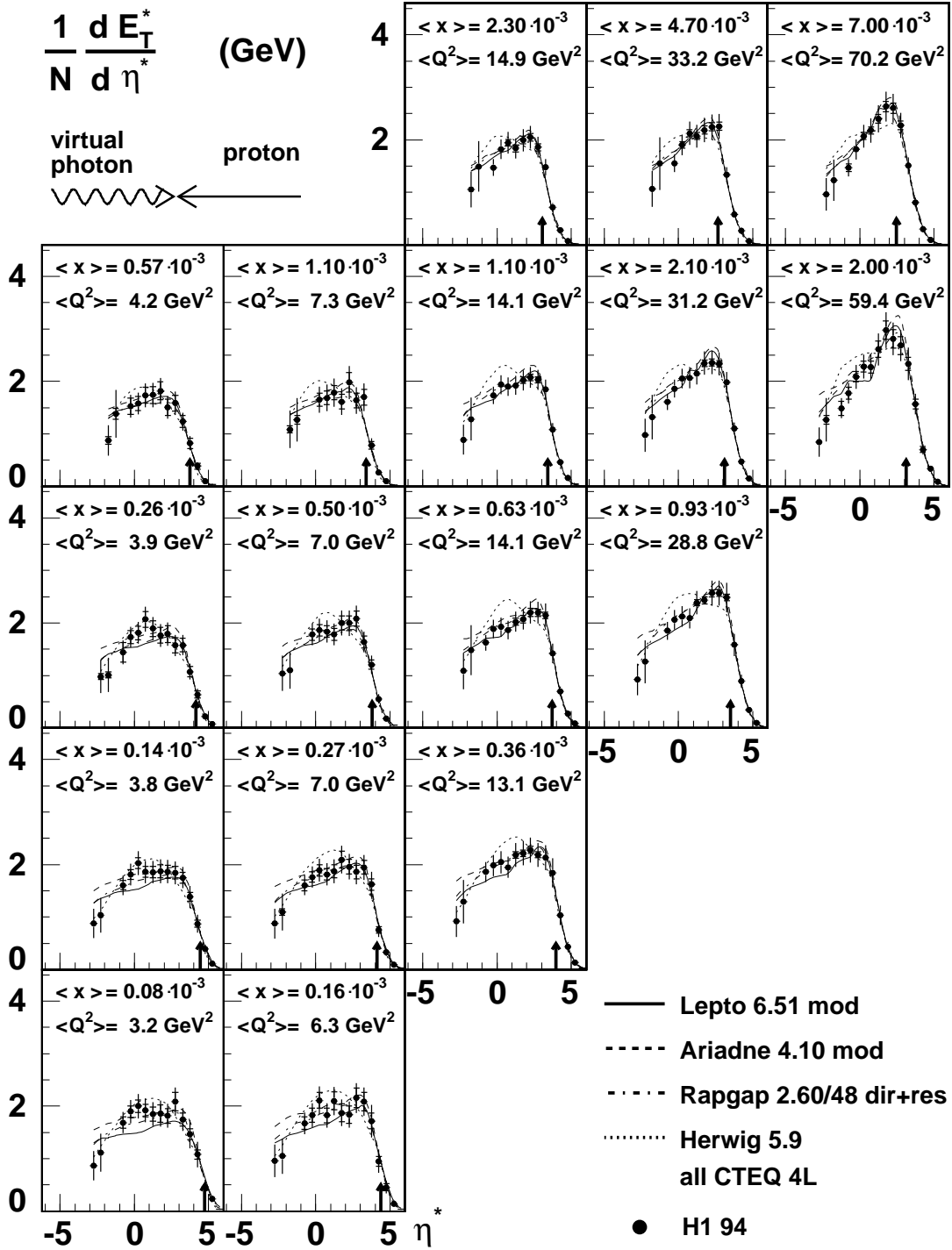,width=14.7cm,clip=}}
  \end{picture}
  \caption{\label{lpl}
  The inclusive transverse energy flow $1/Nd\etd/d\eta^*$
  at different values of $x$ and \qq for the low
   \qq sample. Note that the errors on all of the measurements made at the
two lowest values of $\eta^*$ in each $x$ and \qq interval
are highly correlated and largely independent of the errors
at larger values of $\eta^*$.
% The dominant contributions to the errors on each of the
%data points for $\eta^*<-1$ are totally correlated.
The data are compared to four
  QCD based models.
 The arrows
   mark the average position of the origin of the Breit frame
($\frac{1}{2}\ln{\frac{1}{{\langle}x{\rangle}}-1} $).}
\end{figure}

In Fig.~\ref{lpl} the \etf in the hCMS  is presented in 17 regions
of low $Q^2$ and $x$.
The values  of average \qq 
 cover a range from $3.2$ GeV$^2$ to $70$ GeV$^2$ 
and the mean values of $x$ extend 
 from  $0.08\cdot 10^{-3}$ to $ 7\cdot 10^{-3}$. 
Here, and in all following figures
in which
two sets of error bars are displayed, 
the total errors (outer error bars) are the result of  
adding in quadrature
the statistical errors (inner error bars) and the
systematic errors.

The data exhibit a mean transverse energy of approximately $2$ GeV per unit of
pseudorapidity. The \etf shows a plateau for \qq values below about 
$10$ GeV$^2$ in the current region ($\eta^* > 0$), although as 
\qq increases the distributions become peaked in this region. Using the  
PLUG calorimeter it is possible  to measure \te in the vicinity of 
the proton remnant; these data can be seen as the two  points at smallest
$\eta^*$.   
Here 
the \te flow tends to be about one half of  
 that which is measured in the current region.

%For each $x$ and \qq interval
The systematic errors on the \etd measurements shown in Fig.~\ref{lpl} are
correlated. 
The measurements in the range $-1<\eta^*<-4$ were made predominantly using 
the LAr calorimeter and  suffer from a 
10\% normalisation uncertainty owing to the LAr hadronic energy
 scale and the model dependence of the correction procedure. These 
 are uncorrelated with the errors on the measurements made using  the
PLUG and the BEMC calorimeters. For each $x$ and $Q^2$ interval 
the PLUG was used exclusively for  
  the data points at the two 
lowest values of $\eta^*$ and the BEMC was the main calorimeter used for 
measurements in three highest $\eta^*$ bins.  The errors on the measurements 
provided by both of these calorimeters are  predominantly normalisation uncertainties.

In Fig.~\ref{lpl} four QCD
 models, 
{\sc Ariadne}, {\sc Herwig}, {\sc Lepto} and {\sc Rapgap}  
are compared to the data. 
These models give an acceptable overall 
description of the measured  transverse energy flow with the exceptions  
that  {\sc Herwig} fails to describe the
shape of the distribution for values of $Q^2$ above about 7 GeV$^2$ and
shows a peak
at $\eta^* \approx 1$ and {\sc Lepto} fails to match the data 
in the central pseudorapidity region for the lowest values of 
$x$ and $Q^2$.
{\sc Ariadne} and {\sc Rapgap} give the best description of 
all of the models in the lowest
$x$ and \qq bins although each tends to overestimate the transverse 
energy flow in the vicinity of the proton remnant.
%%Conversely, {\sc Herwig} is in agreement with the data in this
%%region of $\eta^*$. 
% but do not show a
%sufficient suppression of the \te production in the vicinity of the proton. 
%%The
%%largest discrepancies occur in the region of small $x$ and \qq.  

\begin{figure}[p]\unitlength1cm           
   \begin{picture}(16,20)                                
      \put(-0.1,0){\epsfig{file=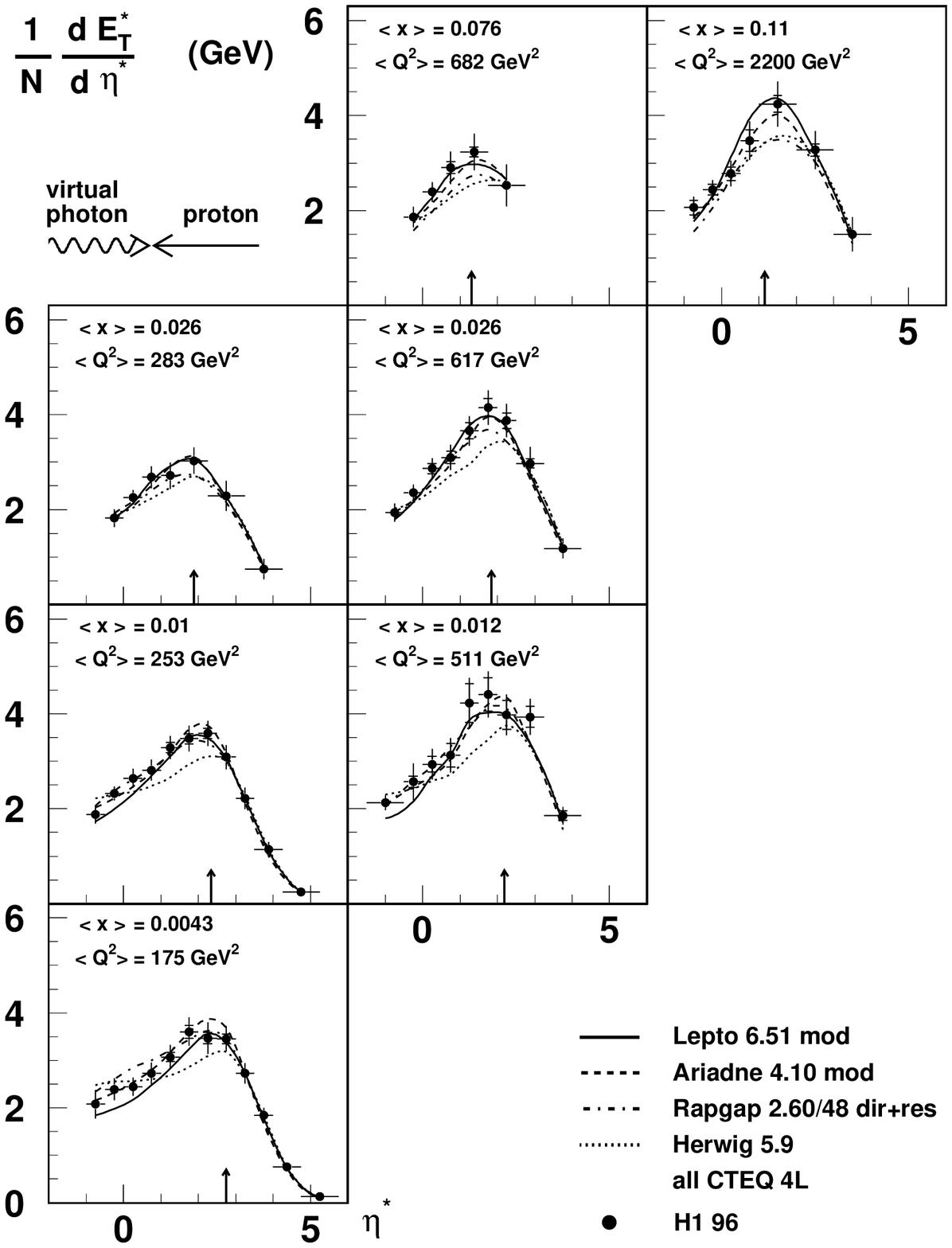,width=16cm,clip=}}
%      \put(10.9,5.){\epsfig{file=gpl_one.eps,width=5.6cm,height=5.89cm,clip=}}
   \end{picture}
   \caption{\label{gpl}
   The inclusive transverse energy flow $1/Nd\etd/d\eta^*$
   at different values of $x$ and \qq for the high 
   \qq sample. The data are compared to four QCD based models. The arrows
   mark the average position of the origin of the Breit frame
($\frac{1}{2}\ln{\frac{1}{{\langle}x{\rangle}}-1} $).
      }
\end{figure}

\begin{figure}[!ht]\unitlength1cm           
   \begin{picture}(16,7.5)                                
      \put(-0.2,-.1){\epsfig{file=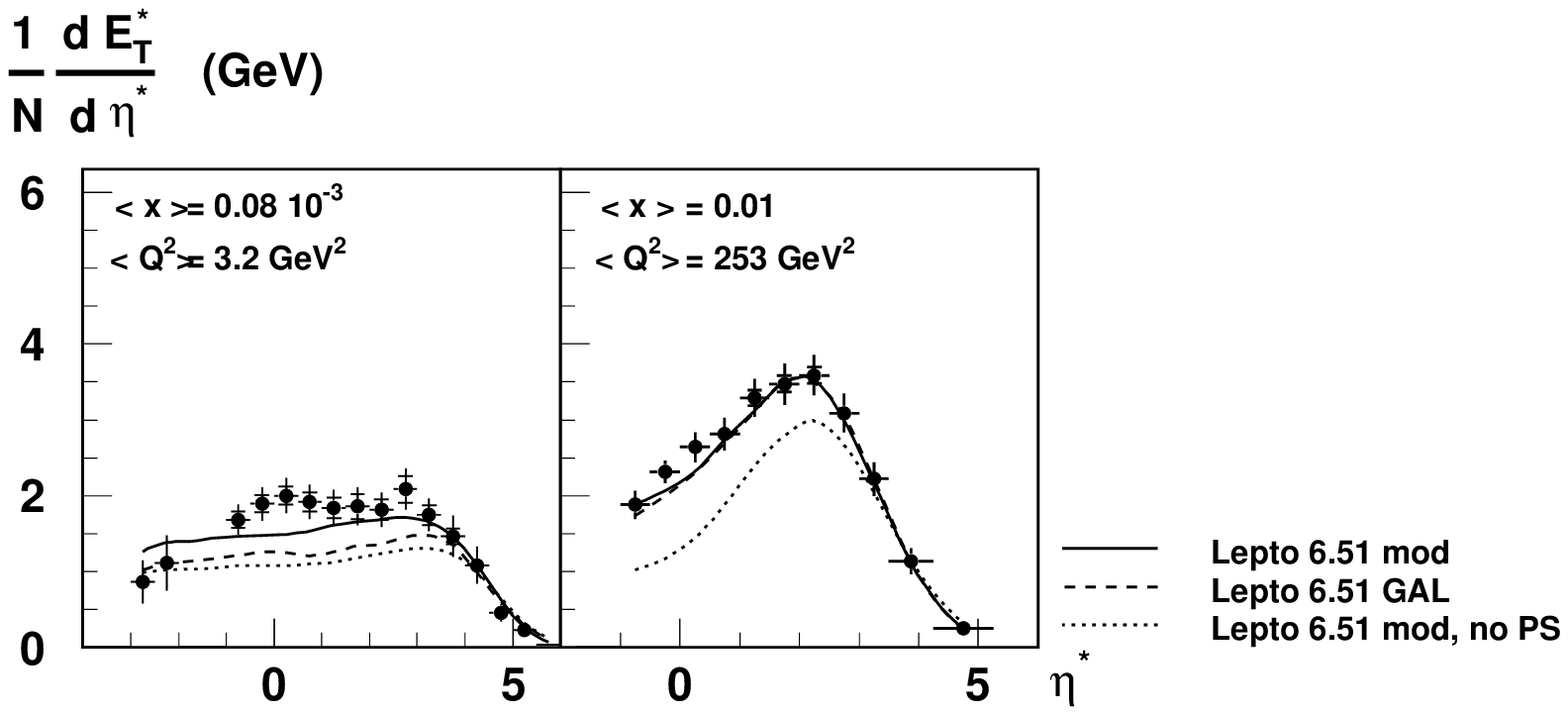,width=15cm,clip=}}
   \end{picture}
   \caption{\label{gone} 
The inclusive transverse energy flow $1/Nd\etd/d\eta^*$  
for two selected kinematic bins from Figs. \ref{lpl} 
and ~\ref{gpl}. 
The influence of GAL string reinteractions and parton showers 
on the  expected  transverse energy flow are shown.
      }
\end{figure}

The measurements at high \qq are shown in Fig.~\ref{gpl}. 
The \etf is presented in 7 regions
in $x$ and \qq and compared again to the four QCD models.
At high \qq the experimental resolution
in $\eta^*$ becomes strongly dependent on $x$ and \qq and the $\eta^*$
bin widths are adjusted accordingly.
The  average \qq values cover a range from $175$ GeV$^2$
to $2\,200$
GeV$^2$ and average $x$ from 0.0043 to 0.11. For 
consistency with the low \qq results the $E_{fwd}$ cut is applied, 
although its
influence is small at high $Q^2$. 
  As is the case for the low \qq spectra,  the \te flow is seen to  peak
in the
current region. Here, the measured  \te  is significantly higher  than
at low
\qq. 
In this kinematic range all QCD models describe the data well 
%(see section
%\ref{QCDpre} for the necessary changes to the models) 
 with the 
exception of {\sc Herwig} which
produces insufficient \te over the \qq range shown  
%In addition {\sc Herwig} 
%can 
and does not 
%well 
describe the shape of the energy flows.

At high \qq the positions of the maxima of the \etf coincide
approximately 
with the average origin of the Breit frame at $\eta^*_{O-BF}=
\frac{1}{2}\ln{\frac{1}{{\langle}x{\rangle}}-1}$. The
Breit frame (BF) is defined as the 
%system where 
frame in which
the exchanged boson carries the
momentum $(0,0,0,-Q)$. 
%%It has been shown~\cite{bfhz} that the hadronic 
%%properties of the BF hemisphere 
%of the 
%%containing the outgoing
%%struck quark resembles those of one half of an $e^+e^-$ interaction. 
%In~\cite{OBF} the
%special role of the BF has been emphasized by the argument that the point of
%the highest perturbative radiation asymptotically coincide with the
%average origin of
%this reference frame. 
%Such a 
In \cite{OBF} it was argued that at high \qq the maximum of the radiated  
transverse energy should coincide with $\eta^*_{O-BF}$.
This behaviour can be seen in our data. In
Fig.~\ref{lpl} and Fig.~\ref{gpl} the position of $\eta^*_{O-BF}$
is marked by an arrow.

In Fig.~\ref{gone} the transverse energy flow is shown for 
two \qq bins (\mean{Q^2}$=3.2$ GeV$^2$ and \mean{Q^2}$=253$ GeV$^2$)  
compared to the predictions of {\sc Lepto} with different parameter 
settings. The effects of using  GAL based string reinteractions   
are shown as
the dashed
line. 
The solid line represents {\sc Lepto} predictions when the GAL
model is not used. The
GAL approach 
leads to a deficit of transverse energy production at low \qq 
although its effect is small at high $Q^2$. The expectations of 
{\sc Lepto} without parton showers (PS) are shown as the dotted curves. 
The predictions lie below the data for most of the pseudorapidity 
range measured, illustrating the sensitivity of our measurements to
 pQCD processes, as  modelled here using a parton shower algorithm.
%This is investigated in more detail in the following section.

\subsection{Dependence on  Bjorken-{\boldmath $x$} }

It has so far proved impossible to distinguish between 
the different QCD evolution schemes mentioned in section \ref{QCDpre} 
using inclusive 
 structure function measurements. It  has
been suggested that the hadronic final state  may prove to be
a more sensitive testing  ground~\cite{mueller}. Measurements of energy
flow~\cite{mk2}, jets and leading particles~\cite{dave} in
the region near  
the  proton remnant have indicated that DGLAP  parton evolution fails to
produce sufficient QCD radiation at low $x$.  On the other hand, 
BFKL evolution predicts a  rise of hard parton emissions at
low $x$ in the central pseudorapidity region of the hCMS. 
%CCFM evolution, which should interpolate smoothly between the DGLAP
%and BFKL regime, 
%is expected to give a similar increase.

\begin{figure}[t]\unitlength1cm
  \begin{picture}(16.,20.5)
    \put(0,0){\epsfig{file=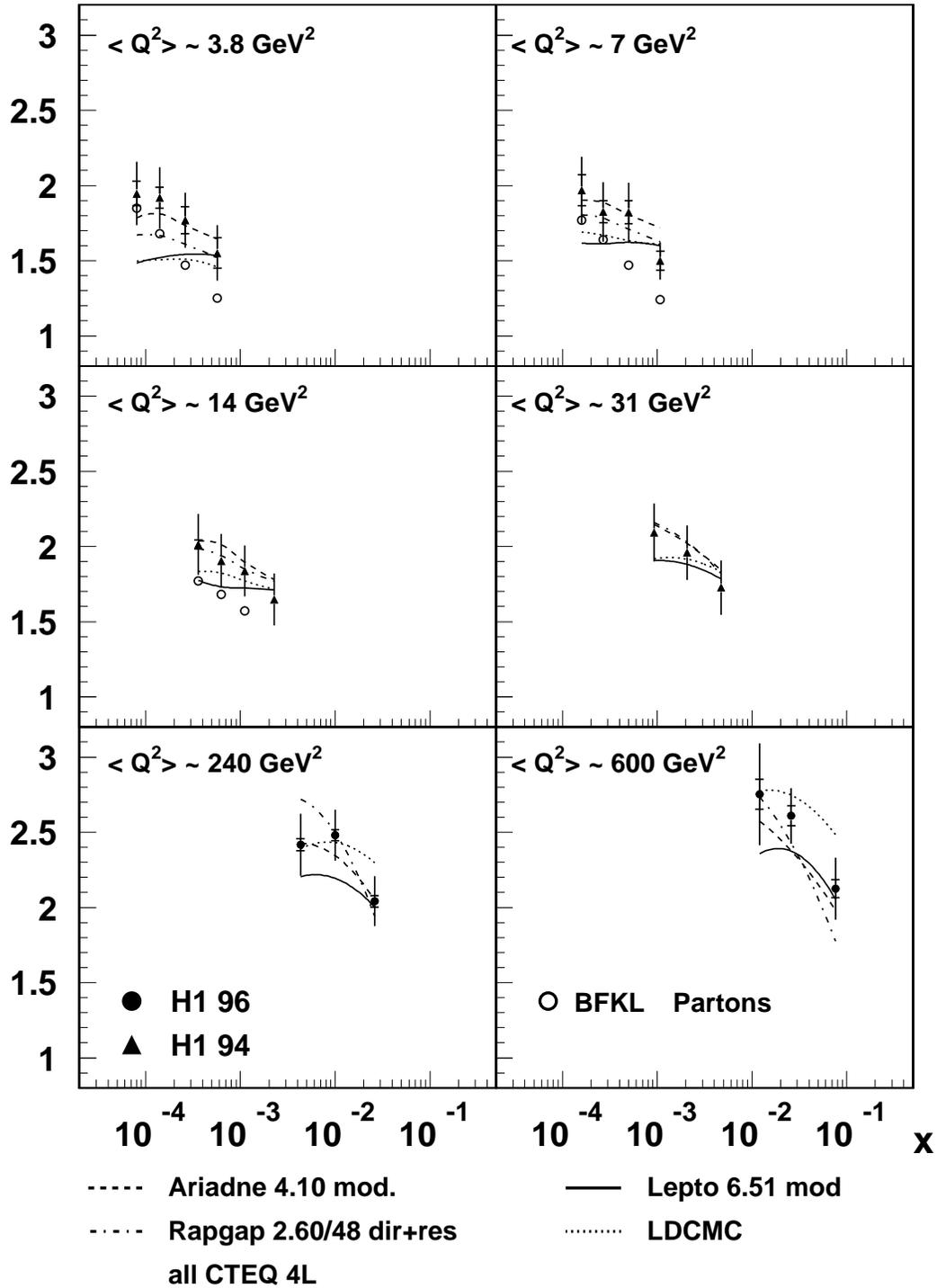,width=14.25cm,clip=}}
  \end{picture}
  \caption{\label{etx}
   Variation of mean $E_T^*$ in the central pseudorapidity region 
($-0.5<\eta^*<0.5$) with $x$ in different regions of \qq 
compared to four QCD based models and an analytical BFKL calculation.}
\end{figure}

To study these QCD evolution effects, 
the dependence of the mean \te in the central 
pseudorapidity region ($-0.5< \eta^* < 0.5$) with $x$ is shown in
Fig.~\ref{etx} 
for different ranges of $Q^2$. The  central region has been chosen since
it
is less affected by the hard scattering 
%(see also section \ref{ww}) 
and
still lies  within the acceptance of the H1 detector. A  rise
in the measured \te is observed as $x$ decreases.  The general behaviour
of the data can be understood as a rise of the average transverse energy
with hadronic mass, $W$, which increases as $x$ decreases at fixed $Q^2$. 

The QCD based Monte Carlo models exhibit reasonable agreement with the
data over most of the kinematic range presented here.  
However, the shape of the $x$ dependence as predicted by 
the DGLAP based model {\sc
Lepto} does not follow closely the data at the smallest
\qq and $x$ values shown in Fig.~\ref{etx}. The calculations of the
CCFM based 
LDCMC show the same behaviour in this region. 
%The idea that the DGLAP approach is insufficient in this range of $x$
% and $\eta$ 
%is a possible explanation 
%and  is supported by the failure $x$ dependence.  
Analytical BFKL calculations at the parton
level~\cite{etbfkl} describe the $x$
dependence better and are closer to the data at the lowest 
values of \qq and $x$.  
However, uncertainties due to hadronisation corrections and  
missing higher orders in the calculations preclude   
an interpretation of the data as a
signal of the onset of BFKL dynamics.

%Furthermore, CCFM as represented by {LDCMC\-} almost coincide 
%with {\sc Lepto} and does not show any rise in the \te with decreasing
%$x$
%at the smallest \qq values, either.

The effects of a resolved component of the virtual photon provide another possible explanation
of the observed increase of \te production with decreasing $x$.
Calculations  using {\sc Rapgap} including  resolved and direct virtual
photon
components  describe well the $x$ dependence and the amount of  
\te flow. The CDM approach of {\sc Ariadne} is also  
in good agreement with the data.  

\subsection{Dependence on  {\boldmath $Q^2$} and {\boldmath
$W$}}\label{ww}

\begin{figure}[t]\unitlength1cm
   \begin{picture}(16,9.9)
      \put(0,0){\epsfig{file=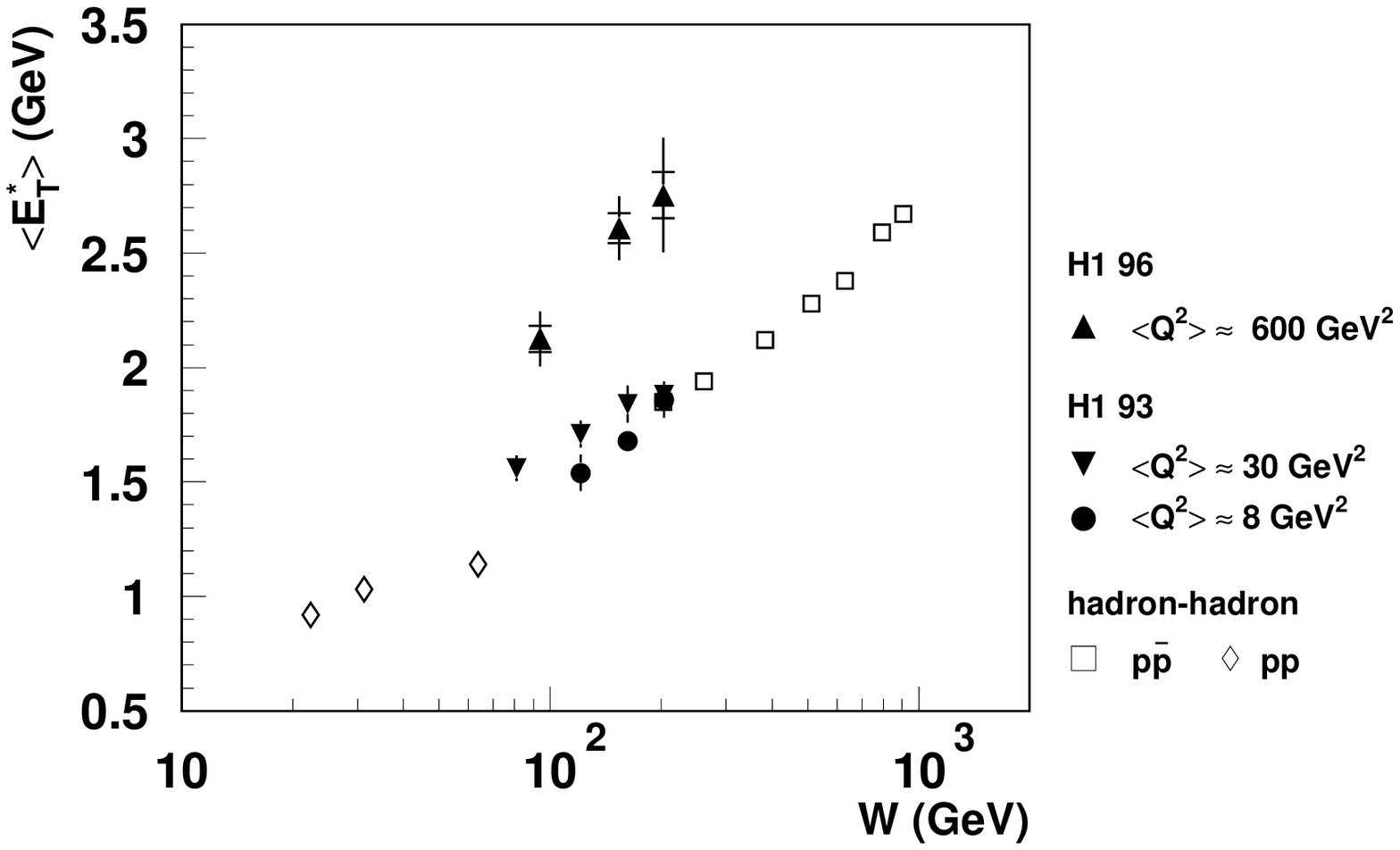,width=16.cm,clip=}}
   \end{picture}
   \caption{\label{wdep}
   Variation of the average $E_T^*$ in the central pseudorapidity region
($-0.5<\eta^*<0.5$) 
   with $W$. The present high \qq data are shown together with earlier H1 
   results at lower $Q^2$. Also shown are  
    hadron-hadron results  
   ($p\bar{p}$ and $pp$ from UA1, NA22 and AFS). 
   } 
\end{figure}

\begin{figure}[ht]\unitlength1cm           
   \begin{picture}(15.5,17.)                                
      \put(-0.15,0){\epsfig{file=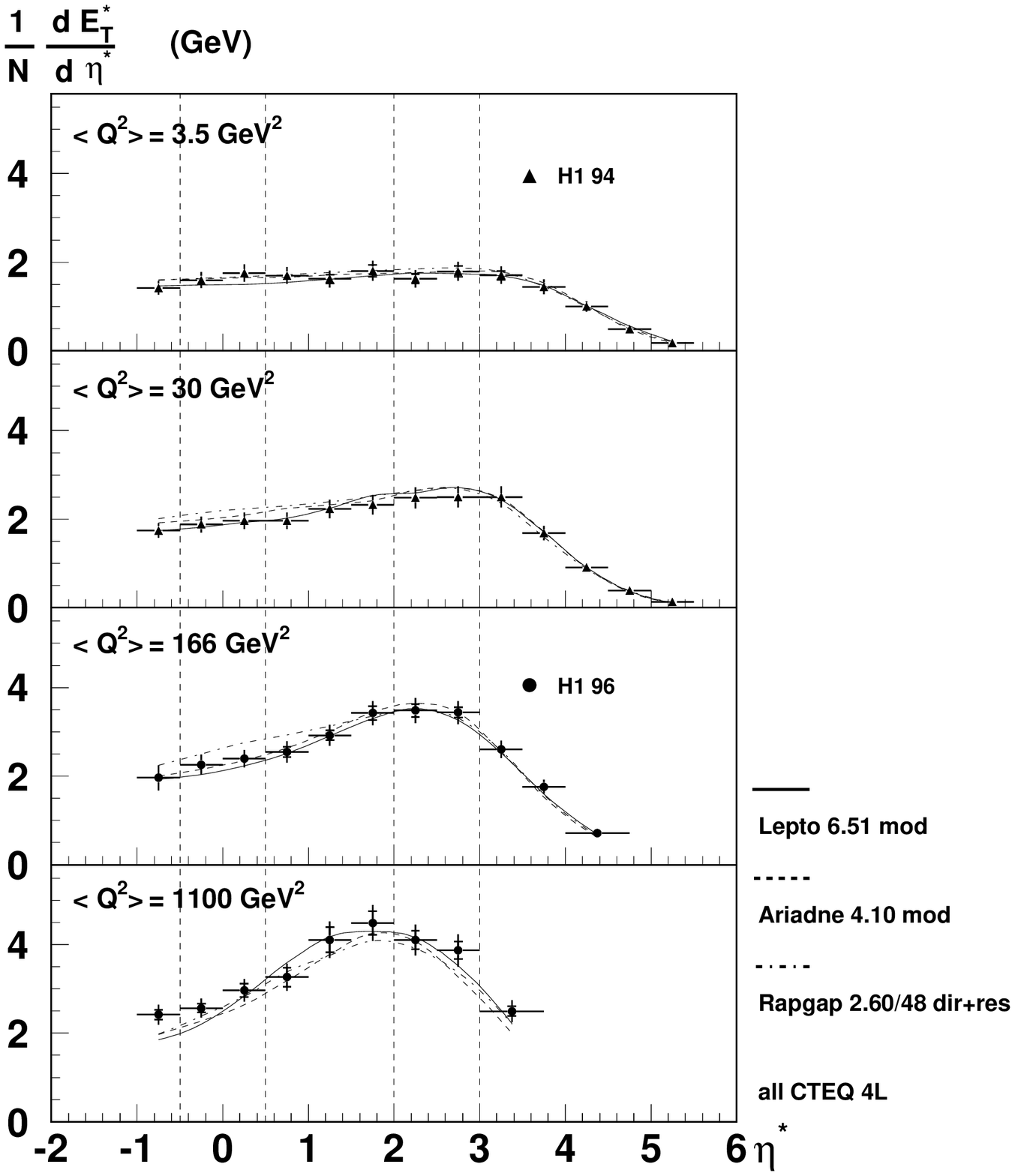,width=15.5cm,clip=}}  
   \end{picture}
   \caption{\label{wpl}
   The inclusive transverse energy flow 
   $1/Nd\etd/d\eta^*$ in different regions of \qq
    for \mean{W} $\approx$ 185 GeV.
    The data are
   compared to three QCD based models.  The two pairs of dashed
vertical lines 
denote the central ($-0.5<\eta^*<0.5$) and photon fragmentation 
($2<\eta^*<3$) bins.  } 
\end{figure}

As mentioned in the introduction, photon-proton scattering 
is similar to a hadron-hadron process, as it can be described as  
the  fluctuation of a photon into a 
 hadronic system which then scatters off the proton. In a previous
publication
on transverse energy flow, H1 showed that this  picture is not only valid
in  photoproduction
but also in DIS~\cite{andre1}. 
 A similar level of \mean{E_T^*} in the central pseudorapidity region and
a $W$ dependence similar to that observed in 
 hadron-hadron interactions were found.  This is consistent 
 with the Bjorken-Kogut picture in which hadronic 
final state quantities in the central pseudorapidity region are expected
to be insensitive to the nature of the colliding particles~\cite{bjoko}
and
depend only on the centre of mass energy.

\begin{figure}[t]\unitlength1cm
   \begin{picture}(16.,12.25)
      \put(0,0){\epsfig{file=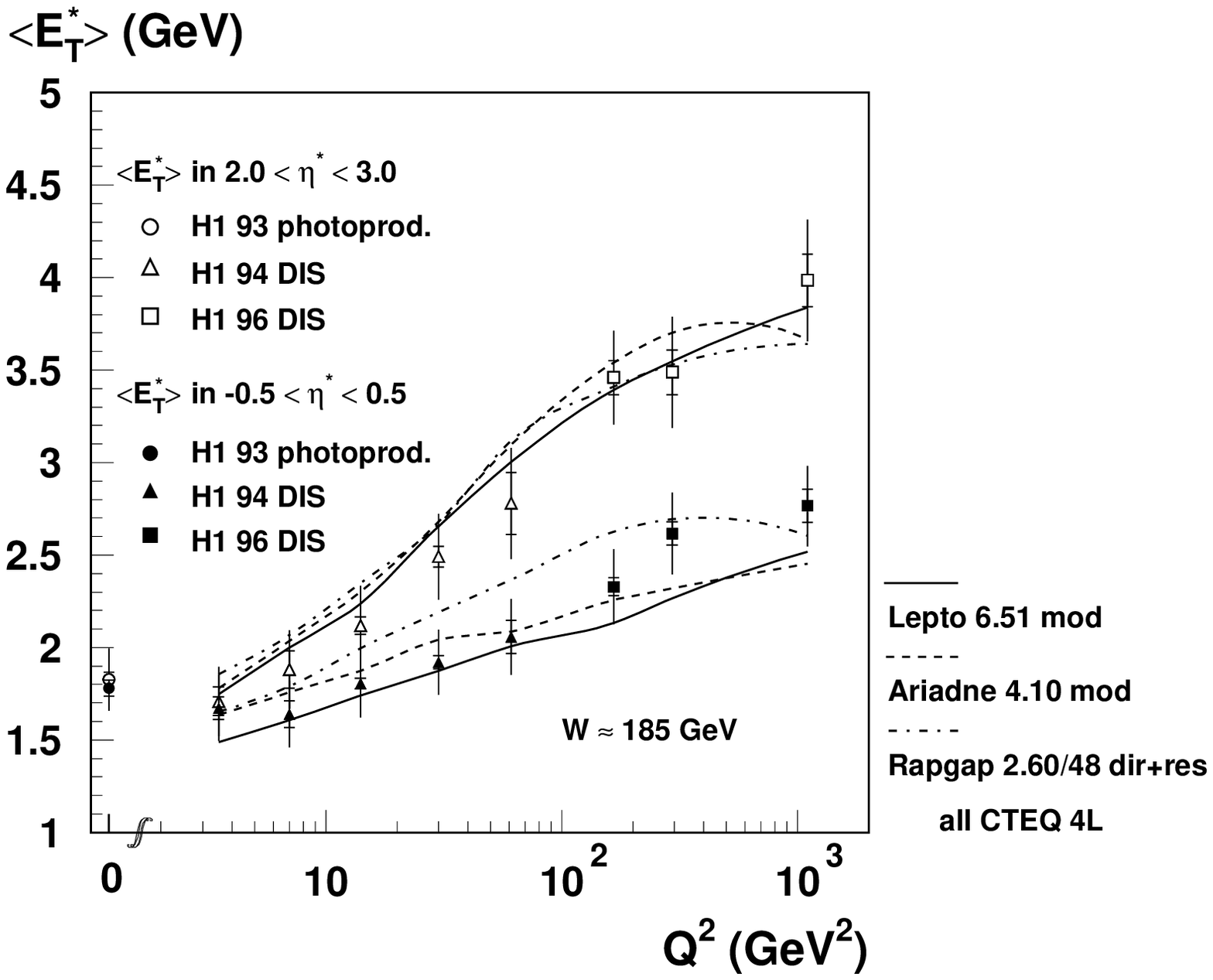,width=16.cm,clip=}}
   \end{picture}
   \caption{\label{www}
   Variation of the mean $E_T^*$ in the central pseudorapidity bin 
($-0.5<\eta^*<0.5$)
   and the photon fragmentation bin ($2<\eta^*<3$) with $Q^2$. The data
   are compared to three QCD based models. 
}
\end{figure}

Fig.~\ref{wdep} shows the $W$ dependence of the new H1 high \qq
data
%\footnote{The 1996 data points in Fig.~\ref{wdep} are taken from
%Fig.~\ref{gpl}. Therefore the $E_{fwd}$ cut is applied but at these high
%\qq
%values the difference is negligible.} 
in the central pseudorapidity region
($|\eta^*|<0.5$) together with our previously published low \qq results. 
Data from $p\bar{p}$ and $pp$ experiments~\cite{ua1,hd}, taken with
non-single 
diffractive triggers,  are also shown. This trigger requirement 
excludes elastic and quasi-elastic scattering events.  
%ensures rejection of hadron-hadron elastic scattering events. 
  
A $W$ dependence compatible with that of the hadron-hadron data 
is found in the H1 data although the mean \te is larger at high
$Q^2$. For a
given value of $Q^2$, the
increase in \mean{E_T^*} seen in the H1 data  is consistent with a linear dependence on $\ln W$.

In order to investigate further the \qq dependence,  data
 have been selected  
from a small range in $W$ ($165$ GeV $<W<$ $213$ GeV). 
 To ensure consistency with the previous study~\cite{andre1},
the $E_{fwd}$ cut is removed 
for this part of the analysis and  data points and model predictions 
 in all subsequent plots are shown with this condition dropped. This has
an effect of, at most, 10\% for 
the lowest \qq points and is negligible for the high \qq measurements. 
These  fixed-$W$ \te flow measurements are shown in 
Fig.~\ref{wpl}. From these distributions, it can be seen that the maximum
of the \etf not only increases with \qq but also that its position
moves towards the central pseudorapidity region.

To study the $\etd$ flow from the fragmenting photon 
and in the central pseudorapidity region, two slices in $\eta^*$ 
are studied, namely the
 central bin ($-0.5<\eta^*<0.5$)  and  the so-called
photon fragmentation  bin ($2<\eta^*<3$). These are delineated in Fig.~\ref{wpl}. 
The definition of the  photon fragmentation bin is somewhat arbitrary.
The chosen
bin in $\eta^*$ is expected to be dominated by the fragmentation of the 
hadronic fluctuation of the photon.  Since
the rapidity range within which particles of mass $m$ can be produced is
limited to $\pm \ln W/m$, and the width of the photon fragmentation 
peak is $\sim \ln (Q^2/m^2)$~\cite{bjoko},  the
peak position of the \te
flow associated with
the fragmenting photon moves towards the central pseudorapidity bin 
 as \qq increases for fixed $W$. This can be seen in Fig.~\ref{wpl}.

Fig.~\ref{www} shows that the measured \te in the photon fragmentation bin
rises with $Q^2$. Qualitatively,
this is consistent with the Generalised Vector Dominance 
Model~\cite{gvdm,sak,gor}, in 
which mesons with higher masses  can 
contribute to the scattering at higher $Q^2$. A more 
formal approach~\cite{wave} has shown that the photon wave function allows
for $q\bar{q}$ states with higher $p_T$ at higher $Q^2$.  

Fig.~\ref{www} also provides the first experimental evidence for a 
rise in the transverse energy with $Q^2$ in the central pseudorapidity
bin. This result was not observed in  previous work~\cite{andre1},
which used less than one tenth of the data available for this analysis 
and which showed no significant $Q^2$ dependence in this region.   

%%%%To investigate the significance 
%%%%of this rise, a 
%first order polynomial 
%%%%straight line
%%%%was fitted to the  
%%%%%distribution of centrally produced \te flow as a function of $\ln  Q^2$.
%%%%%A slope of $\ 0.2\pm0.01$(stat.)$\pm0.1$(syst.) was obtained. 
%%%%%The influence of correlations between the systematic errors on the fit  
%%%%%was investigated. The systematic error due to the LAr energy scale, which  
%%%%%%affects most points equally, was removed. Other normalisation
%%%%%%uncertainties due to model dependence and the calorimeter electromagnetic  
%%%%%%energy scales, which affect separately the  points made with the low
%%%%% and high \qq  data sets, were also studied. The systematic error 
%%%%%on the slope arises from the variation of the low and high \qq points 
%%%%after considering these normalisation uncertainties. 
%%%%%The result of the fit favours a rise with \qq  
%%%%%even
%%%%%although $x$ is still 
%%%%%%%%of the order of $10^{-2}$ even in the highest \qq bin.  

%%The main features of our data can be understood assuming a photon
%%fragmentation
%%region of width $\sim \ln (Q^2/m^2)$, where $m$ is a typical hadron
%%mass.
%%Since
%%the rapidity is limited to $\pm \ln (W/m)$ the photon fragmentation
%%region
%%begins 
%%to overlap with the central rapidity region of the hCMS as \qq increases
%%and this was observed in Fig.~\ref{wpl}. 

  Comparisons are made with {\sc Ariadne}, {\sc Lepto} and {\sc Rapgap}. 
As they are used here, {\sc Lepto} and {\sc Rapgap} 
do not include diffractive events. Therefore, the predictions of 
these models are expected to lie above  
the data  in
Fig.~\ref{www}.  
However, 
%assuming a  
as the 
fraction of diffractive  events 
%of
is
less 
than $10$\%,  with an average transverse energy production of about one
half  
of that observed in non-diffractive interactions~\cite{difET}, 
 the bias is not larger than $10$\%. A comparison 
of {\sc Rapgap} predictions including pomeron exchange processes
with those {\sc Rapgap} calculations shown here also supports this.
All of the QCD based models describe the rise with \qq  
of the measured average transverse energy in the central and
photon fragmentation regions although {\sc Rapgap}  
predicts too much  transverse energy in the central 
pseudorapidity range.

\section*{Summary} 

Measurements of energy production transverse to the photon and 
proton directions in the
hadronic centre of mass
system have been presented using deep-inelastic scattering data 
taken in positron-proton collisions by the H1 Collaboration at HERA. The
measurements cover  more than 8 units of pseudorapidity and extend over
the kinematic range $3.2 < Q^2< 2\,200$ GeV$^2$, $x> 8 \cdot
10^{-5}$ and $66<W<233$ GeV. 

For \qq values below about 10 GeV$^2$ the inclusive 
transverse energy flow $1/Nd\etd/d\eta^*$ shows a plateau-like behaviour 
in the current hemisphere with typical values of about 2 GeV. 
As \qq increases, the distribution shows a peak of increasing 
magnitude.   
For values of \qq  greater than approximately 70 GeV$^2$, the 
peak position  coincides with the origin of the Breit frame. Transverse
energy flow    
in the vicinity of the proton remnant is observed to be 
approximately half of that in the current hemisphere, albeit 
with large systematic uncertainties.

Transverse energy production in the 
central pseudorapidity region rises with increasing $W$ which is 
consistent with observations made in  
 hadron-hadron experiments. However, for the first time 
there 
is evidence of an increase in the level of transverse energy in the 
central rapidity region with $Q^2$. 
%This is at variance with the 
%simple picture of the photon fluctuating into a hadron which 
%then interacts with the proton.

Four QCD based models have been compared to the data. 
  At low $x$ and  $Q^2$,  
predictions made using approaches based on DGLAP and CCFM evolution, 
implemented within  the {\sc Lepto}
and {\sc LDCMC} models, respectively,  are not able to 
fully describe  the transverse
energy flow in the central
pseudorapidity region.  {\sc Rapgap}, which includes resolved
virtual photon processes, and the Colour Dipole Model, as implemented
in {\sc Ariadne}, give a reasonable description of the data 
in this region. Furthermore, the {\sc Lepto} model can only 
describe the measurements over the full current hemisphere at low \qq
 if a General Area Law based string reinteractions scheme is not used.  At
the highest
\qq 
values, all models describe the data well except for {\sc Herwig} which 
provides a relatively poor description over most of the kinematic range
considered in this analysis.

\section*{Acknowledgements}
We are grateful to the HERA machine group whose outstanding efforts have made
and continue to make this experiment possible. We thank the engineers and
technicians for their work constructing and maintaining the H1 detector, our
funding agencies for financial support, the DESY technical staff for continual
assistance and the DESY directorate for the hospitality which they extend 
to the non-DESY members of the collaboration.


\begin{thebibliography}{99}
\bibitem{ua1} UA1 Collaboration, {C.~Albajar et al.},
\Journal{\NPB}{335}{261}{1990}.
\bibitem{charged} H1 Collaboration, {I.~Abt et al.},
\Journal{\ZPC}{63}{377}{1994}.
%
%Nucl. Phys. B485 (1997) 3.
%\bibitem{
%\bibitem{mk0} Michael Kuhlen, QCD and the Hadronic Final State in Deep Inelastic 
%Scattering at HERA, Habilitationsschrift Thesis, University of Hamburg, 1997 and referen
\bibitem{mk1} H1 Collaboration, C.~Adloff et al.,
\Journal{\NPB}{485}{3}{1997}.
\bibitem{h1f297} H1 Collaboration, {C.~Adloff et al.}, 
\Journal{\NPB}{497}{3}{1997}.
\bibitem{zeusf297} ZEUS Collaboration, {J.~Breitweg et al.},
\Journal{\EPC}{7}{609}{1999}.
\bibitem{DGLAP}
  Yu.~L.~Dokshitzer, Sov. Phys. JETP 46 (1977) 641; \\
  V.~N.~Gribov and L. N. Lipatov, Sov. J. Nucl. Phys. 15 (1972) 438 and
675; \\
  G.~Altarelli and G.~Parisi, \Journal{\NP}{126}{297}{1977}. 126 (1977).
\bibitem{bfkl} E.~A.~Kuraev, L.~N.~Lipatov and V.~S.~Fadin,
                Sov. Phys. JETP 45 (1972) 199; \\
                Y.~Y.~Balitsky and L.~N.~Lipatov,
                Sov. J. Nucl. Phys. 28 (1978) 822.
\bibitem{CCFM} M.~Ciafaloni, \Journal{\NPB}{296}{249}{1987}; \\
                S.~Catani, F.~Fiorani and G.~Marchesini,
                \Journal{\PLB}{234}{339}{1990}, \Journal{\NPB}{336}{18}{1990}.

%\bibitem{mk1} H1 Collaboration, {I. ~Abt et al.}, \Journal{\ZPC}{63}{377}{1994}.
\bibitem{mk2} H1 Collaboration, {S.~Aid et al.}, \Journal{\PLB}{356}{118}{1995}.
\bibitem{tania} H1 Collaboration, {C.~Adloff et al.}, \Journal{\PLB}{415}{418}{1997}.
\bibitem{zeusfj} ZEUS Collaboration, {J.~Breitweg et al.}, \Journal{\EPC}{6}{239}{1999}.
\bibitem{dave} H1 Collaboration, {C.~Adloff et al.}, \Journal{\NPB}{538}{3}{1999}.
\bibitem{bjoko}  J.D.~Bjorken and J.B.~Kogut, \Journal{\PRD}{8}{1341}{1973}.
\bibitem{duca}   V.~Del~Duca et al., \Journal{\PRD}{46}{931}{1992}.
\bibitem{shaw} G.~Shaw, \Journal{\PLB}{318}{221}{1993}; \\ 
                G.~Kerley and G.~Shaw, \Journal{\PRD}{56}{7291}{1997}.
\bibitem{ioffe} B.~I.~Ioffe, \Journal{\PLB}{30}{123}{1969}.
\bibitem{andre1} H1 Collaboration, {S.~Aid et al.},
\Journal{\PLB}{358}{412}{1995}.
\bibitem{na22} NA22 Collaboration, {M.~Adamus et al.},
\Journal{\ZPC}{39}{311}{1988}.
\bibitem{h1nim}
H1 Collaboration, { I.~Abt et al.}, \Journal{\NIMA}{386}{310 and
348}{1997}. 
%H1 Collaboration, { I.~Abt et al.}, \Journal{\NIMA}{386}{348}{1997}.
\bibitem{larc} H1 Calorimeter Group, {B.~Andrieu et al.}, \Journal{\NIMA}{336}{460}{1993}.
\bibitem{h1po} H1 Calorimeter Group, {B.~Andrieu et al.}, \Journal{\NIMA}{350}{57}{1994}.
\bibitem{h1pi}
H1 Calorimeter Group, {B.~Andrieu et al.}, \Journal{\NIMA}{336}{499}{1993}.
\bibitem{h1bemc} {J.~Ban et al.}, \Journal{\NIMA}{372}{399}{1996}.  
%\bibitem{SPACAL} H1 SPACAL Group, {R. D. Appuhn et al.},
\bibitem{f2pap} H1 Collaboration, {S. ~Aid et al.}, \Journal{\NPB}{470}{3}{1996}.
%\bibitem{SPACALTEST} H1 SPACAL Group, {R. D. Appuhn et al.},
%\Journal{\NIMA}{386}{397}{1997}.
%
\bibitem{shad} H1 SPACAL Group, {R.D.~Appuhn et al.},
\Journal{\NIMA}{382}{395}{1996} 
%\bibitem{spaca} A.~Meyer, Ph.D. thesis, Hamburg 1997. 
\bibitem{spaca} A.~Meyer, DESY internal report FH1-97-01, 1997.
\bibitem{enrico} E.~Panaro, DESY-THESIS-1998-020, 1998.
%\bibitem{enrico} E.~Panaro, Ph.D thesis, Hamburg 1998.
\bibitem{jablo} F.~Jacquet and A.~Blondel in
                 Proc.\ of the Study of an ep Facility for Europe, ed.
                 U.~Amaldi, DESY-79-48, p.391 (1979).
\bibitem{sigma} U.~Bassler and G.~Bernardi,\Journal{\NIMA}{361}{197}{1995}.
%\bibitem{f2pap} H1 Collaboration, {S.~Aid et al.},
%\Journal{\NPB}{470}{3}{1996}.

\bibitem{DJANGO} G.A.~Schuler and H.~Spiesberger, Proc. of the HERA
Workshop, 
eds W.~Buchm\"uller and G.~Ingelman (1992) Vol. 3, 1419.
\bibitem{fabian} M.F.~Hess,
 Ph.D.\ thesis, Hamburg 1996, MPI-PhE/96-16.

\bibitem{ariadne} {L.~L\"onnblad, \Journal{\CPC}{71}{15}{1992}.}
\bibitem{herw} G.~Marchesini et al., \Journal{\CPC}{67}{465}{1993}.
\bibitem{lepto} G.~Ingelman,
                Proc. of the HERA workshop, eds W.~Buchm\"uller and
                G.~Ingelman, Hamburg (1992) Vol. 3, 1366.
%\bibitem{SCI}  A.~Edin, G.~Ingelman, J.~Rathsman,
%                \Journal{\PLB}{366}{371}{1996}.   
%\bibitem{mcworkshop} A.~Edin, G.~Ingelman, to appear in the ``Proc.\ of
%the Workshop of Monte Carlo Models at HERA" (1999)
%\bibitem{GAL} J.~Rathsman, preprint SLAC-PUB-8034-Rev, hep-ph9812423v2, March 1998.
%
\bibitem{GRV}  M.~Gl\"uck, E.~Reya and A.~Vogt, \Journal{\ZPC}{67}{433}{1995}.
\bibitem{MRSH} A.D.~Martin, R.G.~Roberts and W.J.~Stirling , MRS Parton
                Distributions, Proc.\ of the Workshop on Quantum Field Theory
                Theoretical Aspects of High Energy Physics, eds. B.~Geyser and
                E.M.~Ilgenfritz, p.11 (1993).
\bibitem{geisha} H.~Fesefeldt, The Simulation of Hadronic Showers - Physics
                  and Applications, RWTH Aachen, PITHA 85/02 (1985).
\bibitem{calor}  H.~Bertini, \Journal{\PR}{188}{1711}{1969}.
\bibitem{phojet} {R.~Engel, Proc.\ of the XXIXth Rencontre de Moriond
(1994)
321.}



\bibitem{evs} M.~Dasgupta and B.R.~Webber, \Journal{\EPC}{1}{539}{1998};
\\
              H1 Collaboration, {C.~Adloff et al.},  \Journal{\PLB}{406}{256}{1997}.
\bibitem{mlla} Yu.L.~Dokshitzer, et al., ``Basics of Perturbative QCD",
Editions Frontieres (1991);
\\
 Ya.I.~Azimove, Yu.L.~Dokshitzer, V.A.~Khoze and S.J.~Troyan,
\Journal{\ZPC}{31}{213}{1986}, \Journal{\ZPC}{27}{65}{1985}; \\
 L.V.~Gribov, Yu.L.~Dokshitzer, V.A.~Khoze and S.J.~Troyan,
\Journal{\PLB}{202}{276}{1988}.
\bibitem{lattice} S.R.~Sharpe, Proceedings of 29th International
Conference on High Energy Physics (ICHEP 98), Vancouver, Canada, 
July 1998, hep-lat/9811006.
\bibitem{neweth} P.~Nadolsky, D.~R.~Stump and C.~P.~Yuan, 
             MSUHEP-90601, June 1999, hep-ph/9906280.
\bibitem{CTEQ}   CTEQ Collaboration, \Journal{\PRD}{55}{1280}{1997}.
\bibitem{pdflib} H.~Plothow-Besch, PDFLIB Version 7.09, User's Manual, W5051
                  1997.07.2 CERN-PPE, 1997.
\bibitem{sasgam} G.~Schuler and T.~Sj\"ostrand, \Journal{\PLB}{376}{193}{1996}.

\bibitem{dipole} {B.~Andersson, G.~Gustafson and L.~L\"onnblad,}
\Journal{\NPB}{339}{393}{1990}.
\bibitem{cdmbfkl}
 L.~L\"onnblad, Z. Phys. C65 (1995) 285; \\
 A.~H. Mueller, Nucl. Phys. B415 (1994) 373;\\
 L.~L\"onnblad, Z. Phys. C70 (1996) 107.
\bibitem{ldc}  B.~Andersson, et al., \Journal{\NPB}{463}{217}{1996}; \\ 
                B.~Andersson, et al., \Journal{\ZPC}{71}{613}{1996}.

\bibitem{string}
B.~Andersson et al.,
Phys.\ Rep.\ {97} (1983) 31.
\bibitem{cluster} B.R.~Webber, \Journal{\NPB}{238}{492}{1984}.

%\bibitem{jetset74}
%T. ~Sj\"ostrand, Comp. Phys. Comm.\ { 82} (1994) 74.
%\bibitem{dirkdis} D.~Kr\"ucker in Proc.\ of the 6th International
%Workshop
%on
%DIS
%and QCD, Brussels 1998.
\bibitem{arpr} L.~L\"onnblad, private communication.
\bibitem{tune} N.~Brook et al., in Proc.\ of the Workshop Future 
                Physics at HERA, eds. G.~Ingelman, A.~De~Roeck and
                R.~Klanner, DESY (1996) Vol. 1,  613.
\bibitem{ldcmc} H.~Kharraziha and L.~L\"onnblad, The Linked Dipole Chain Monte
                 Carlo, preprint LU-TH~97-21, NORDITA-97/54 P, hep-ph/9709424,
                 December 1997.
\bibitem{SCI}  A.~Edin, G.~Ingelman, J.~Rathsman,
                \Journal{\PLB}{366}{371}{1996}.
%\bibitem{mcworkshop} S.~Chekanov, G.~Ingelman, D.~Milstead, to appear in
%the ``Proc.\ of the Workshop of Monte Carlo Models at HERA" (1999).
\bibitem{mcworkshop} ZEUS Collaboration, J. ~Breitweg { et
al.}, DESY-99-041, accepted by \Journal{\EPC}{}{}{1999}.
%%\bibitem{GAL} J.~Rathsman, preprint SLAC-PUB-8034-Rev, hep-ph9812423v2,
%%1998.
\bibitem{GAL} J. ~Rathsman, \Journal{\PLB}{452}{364}{1999}.
%\bibitem{chris} H1 Collaboration, (C.~Adloff et al.),
%                \Journal{\PLB}{428}{206}{1998}.
%\bibitem{eddi} H1 Collaboration, (C.~Adloff et al.),
%                \Journal{\EPC}{5}{439}{1998}.
\bibitem{rapgap} {H.~Jung}, \Journal{\CPC}{86}{147}{1995}.\\
(for update see http://www-h1.desy.de/\string~jung/rapgap/rapgap.html).
%(for update see http://www-h1.desy.de/\string ~jung/rapgap/rapgap.html).

\bibitem{jetset74}
T. ~Sj\"ostrand, Comp. Phys. Comm.\ { 82} (1994) 74.

\bibitem{mecor} M.~Seymour, Lund preprint LU-TP-94-12 (1994); \\  
                M.~Seymour, \Journal{\NPB}{436}{443}{1995}.

\bibitem{etbfkl} K.~Golec-Biernat et al.,
                 \Journal{\PLB}{335}{220}{1994} ; \\ 
P.~Sutton private  communication.



\bibitem{h1r}
H1 Collaboration, { C.~Adloff  et al.}, \Journal{\ZPC}{76}{613}{1997}. 
\bibitem{zr}
ZEUS Collaboration, J.~Breitweg { et al.}, \Journal{\EPC}{6}{43}{1999}. 
%\bibitem{bfhz}  ZEUS Collaboration, {M. ~Derrick et al.},
%                 \Journal{\ZPC}{67}{93}{1995}; \\    
%                 H1 Collaboration, {C. Adloff et al.},
%                 \Journal{NPB}{504}{3}{ 1997}.

\bibitem{OBF}   Yu.~L.~Dokshitzer, Plenary talk at Int.\ Workshop on Deep  
                 Inelastic Scattering and Related Subjects, Eilat Feb.~1994
                 (unpublished); \\ 
               V.~A.~Khoze and W.~Ochs,
                 \Journal{\IJMPA}{12}{2949}{1997}.

\bibitem{mueller}
  A.H.~Mueller, Nucl. Phys.  B (Proc. Suppl.)  18C (1990) 125; \\ 
  A.H.~Mueller  J. Phys. G17 (1991) 1443.
\bibitem{hd}    A.~De~Roeck,
                 Inclusive Particle Production in Hadron-Proton
                  Interactions at 250 GeV/c,
                  Ph.D. Thesis, Antwerpen 1988.
\bibitem{gvdm} A.~Donnachie and G.~Shaw, in Electromagnetic Interactions
of
Hadrons, Vol. 2, 169-194, eds. A.~Donnachie and G.~Shaw, Plenum, New York,
1978.

\bibitem{sak} J.J.~Sakurai and D.~Schildknecht,
\Journal{\PLB}{40}{121}{1972}.
\bibitem{gor} B.~Gorczyca and D.~Schildknecht,
\Journal{\PLB}{47}{71}{1973}.

\bibitem{wave} {S.~Brodsky et al.}, \Journal{\PRD}{50}{3134}{1994}; \\ 
                {S.~Brodsky and P.~Lepage}, \Journal{\PRD}{22}{2157}{1980}.
\bibitem{difET}   H1 Collaboration, S.~Aid et al., \Journal{\ZPC}{70}{609}{1996}.



%%%%%%%%%%%%%%%%%**************










%\bibitem{h1ef1} M.F.~Hess,
%Ph.D.\ thesis, Hamburg 1996, MPI-PhE/96-16.
%\bibitem{det}H1 Collaboration (I.Abt et al.), \Journal{\NIMA}{368}{310}{1997}, ibid p.386.
%\bibitem{cluster} B.R. ~Webber, \Journal{\NPB}{238}{492}{1984}

%\bibitem{lund} B.\ Andersson, G.\ Gustafson, G.\ Ingelman and T.\ Sj\"ostrand,
%Phys.\ Rep.\ {\bf 97} (1983) 31.
%and QCD, Brussels 1998.
%\bibitem{ariadne} {L. L\"onnblad, \Journal{\CPC}{71}{15}{1992}.}
%\bibitem{ext}  B.~Anderson et al., \Journal{\ZPC}{43}{625}{1989}.
%\bibitem{expgap} ZEUS Collaboration (M.~Derrick et al.),
%                  \Journal{\PLB}{315}{481}{1993};\\
%                   H1 Collaboration (T.~Ahmed et al.),
%                  \Journal{\NPB}{429}{477}{1994}.
%(for update see http://www-h1.desy.de/\string~jung/rapgap/rapgap.html).
%\bibitem{CCFMcal} G.~Bottazzi, G.~Marchesini, G.P.~Salam and M.~Scorletti,
%                hep-ph/9810546, IFUM-634-FT; G.P.~Salam private communication.
\end{thebibliography}
\end{document}